\documentclass[pre, reprint, showpacs, superscriptaddress]{revtex4-1}
\usepackage{amsmath, graphicx,amssymb,color, bm}
\usepackage{balance} 
\usepackage{sistyle}
\usepackage[utf8]{inputenc}
\usepackage[T1]{fontenc}
\usepackage[english]{babel}

\newcommand\myeq{\mathrel{\overset{\makebox[0pt]{\mbox{\normalfont\tiny\sffamily def}}}{=}}}

\begin{document}

\title{Yield stress and elasticity influence on surface tension measurements}

\author{Loren J\o rgensen}
\email{loren.jorgensen@univ-lyon1.fr}
\affiliation{Institut Lumi\`ere Mati\`ere, UMR5306 Universit\'e Claude Bernard Lyon~1~- CNRS, Universit\'e de Lyon, 69622 Villeurbanne, France}
\author{Marie Le Merrer} 
\affiliation{Institut Lumi\`ere Mati\`ere, UMR5306 Universit\'e Claude Bernard Lyon~1~- CNRS, Universit\'e de Lyon, 69622 Villeurbanne, France}
\author{H\'el\`ene Delano\"e-Ayari} 
\affiliation{Institut Lumi\`ere Mati\`ere, UMR5306 Universit\'e Claude Bernard Lyon~1~- CNRS, Universit\'e de Lyon, 69622 Villeurbanne, France}
\author{Catherine Barentin} 
\affiliation{Institut Lumi\`ere Mati\`ere, UMR5306 Universit\'e Claude Bernard Lyon~1~- CNRS, Universit\'e de Lyon, 69622 Villeurbanne, France}
\affiliation{Institut Universitaire de France}

\begin{abstract}
We have performed surface tension measurements on carbopol gels of different concentrations and yield stresses. Our setup, based on the force exerted by a capillary bridge on two parallel plates, allows to measure an effective surface tension of the complex fluid and to investigate the influence of flow history. More precisely the effective surface tension measured after stretching the bridge is always higher than after compressing it. The difference between the two values is due to the existence of a yield stress in the fluid. The experimental observations are successfully reproduced with a simple elasto-plastic model. The shape of successive stretching-compression cycles can be described by taking into account the yield stress and the elasticity of the gel. We show that the surface tension $\gamma_{LV}$ of yield stress fluids is the mean of the effective surface tension values only if the elastic modulus is high compared to the yield stress. This work highlights that thermodynamical quantities measurements are challenged by the fluid out-of-equilibrium state implied by jamming, even at small scales where the shape of the bridge is driven by surface energy. Therefore setups allowing deformation in opposite directions are relevant for measurements on yield stress fluids.
\end{abstract}

\maketitle

\section{Introduction}
Yield stress fluids are widespread materials in everyday life, food industry, cosmetics, building industry, oil industry and many other fields. They are of a great interest because they have the property to flow only when the applied stress is greater than a critical stress called yield stress \cite{larson_structure_1999}. They include emulsions, suspensions, gels, granular pastes and foams.

Their bulk properties have been studied extensively since the work of Herschel and Bulkley in 1926 \cite{herschel_konsistenzmessungen_1926} and are now well characterized \cite{coussot_yield_2014}. Besides, capillarity and wetting are well known for simple fluids. Recently a lot of work has also been done on surface tension and wetting of soft solids \cite{mora_capillarity_2010,salez_adhesion_2013,weijs_capillarity_2014} and on the competition between capillary forces and elasticity of the substrate \cite{jerison_deformation_2011,marchand_capillary_2012}. But until now few studies have focused specifically on the surface tension of yield stress fluids \cite{boujlel_measuring_2013}.

However surface tension and wetting properties of yield stress fluids are of a great importance for capillary imbibition, coating, surface instabilities and adhesion, among other applications.

Here we explore the competition between surface tension, which is an equilibrium property to be measured, and yield stress effects that often keep the system out of thermodynamical equilibrium due to a dynamical arrest of flow. This situation can be compared to contact angle hysteresis: the contact angle is always smaller or greater than the equilibrium (Young) value, depending on the history of the contact line \cite{de_gennes_capillarity_2003}.

G\'eraud \emph{et al.} studied this competition in capillary rise experiments \cite{geraud_capillary_2014}; this method allowed to measure the surface tension and the yield stress of the fluid at the same time. Yet, with their setup, a large amount of liquid is needed, the contact angle must be measured in another experiment, the results are extremely sentitive to the least defect on the inner surface of the capillaries and only fluids of low yield stress ($<20$ Pa) can be characterized this way.

The method presented here allows to get rid of these difficulties. Moreover both extension and compression of the system can be imposed, which highlights the effect of the flow history on the effective surface tension measured.

The article is built as follows. In the first part, we present the fluids on which we performed measurements and the experimental setup. Then we describe the experimental results and in the next part we compare them to an elastoplastic model using only few ingredients. Finally we discuss the agreement between the experiments, the model and results from other works.

\section{Materials and methods}

\subsection{Fluids}
The simple fluids used here are deionized water (18 M$\ohm$) and silicon oil (47V100 from Roth).
The complex fluids are carbopol gels of different concentrations. The raw polymer (powder) is ETD 2050 from 
Lubrizol. The gel is prepared as follows: a small amount of polymer is weighted and slowly dissolved in deionized 
water heated at 50$\degC$ and stirred. The hot solution is stirred for 30 minutes, then it is let to cool down to room 
temperature. Evaporation is hindered by covering the container with Parafilm. Sodium hydroxyde (10M) is added to 
the solution until its pH is raised to $7\pm0.5$, which causes the polymer chains to charge 
negatively. The charged chains thus repel each other, the polymer blobs swell and jam, and the solution becomes a 
gel. Finally the gel is either stirred gently by hand or stirred for 24 hours at 2100 rpm with a mechanic 
stirrer. It was indeed shown in other works \cite{baudonnet_effect_2004} that stirring changes the rheology of carbopol. Our carbopol concentrations range from 0.25\% (in weight) to 2\%. Hand stirred (respectively machine stirred) carbopol is denoted HS (respectively MS) in the following.

\subsection{Rheology}
Carbopol gels are generally considered as model, non thixotropic, yield stress fluids. As long as slip \cite{meeker_slip_2004}, transient shear banding \cite{divoux_stress-induced_2011} and confinement \cite{geraud_confined_2013} are avoided, their flow curve, relating the shear stress $\sigma$ to the shear rate $\dot{\gamma}$, is well fitted with a Herschel-Bulkley (HB) law:
\begin{equation*}
\bigg\{
  \begin{tabular}{lcl}
   $\dot{\gamma} = 0$ & \ & if $ \sigma < \sigma_Y $ \\
  $\sigma = \sigma_Y + K \dot{\gamma}^{\ n}$ & \ & if $\sigma \ge \sigma_Y $  
  \end{tabular}
\end{equation*}
with $\sigma_Y$ the yield stress, $K$ the consistency and $n$ the HB exponent.

Our rheometer is an Anton Paar MCR 301 equipped with a rough cone and plate geometry of angle $4^\circ$. The flow curve is obtained with decreasing steps of constant shear rate, ranging from 100 s$^{-1}$ to 0.01 s$^{-1}$ (10 points per decade). The duration of each step is set automatically by the rheometer (between 15~s and 30~s per step) and the measurement is made when the steady state is reached. The elastic modulus $G'$ is measured by oscillatory shear deformation of 1\% with increasing, then decreasing frequencies, ranging from 0.1~Hz to 50~Hz.

The yield stress of our carbopol samples ranges from $\sigma_Y= 0.3 $~Pa to $\sigma_Y= 38 $~Pa and their elastic modulus at 0.1~Hz from $G'= 1.5$~Pa to $G'= 155 $~Pa. Typically $n$ is always between 0.5 and 0.6 and $K$ ranges from 0.75~Pa.s$^n$ to 13~Pa.s$^n$.

\subsection{Bridge tensiometer}
\subsubsection{Setup.~~}
\begin{figure}[floatfix,h]
\centering
  \includegraphics[width=\columnwidth]{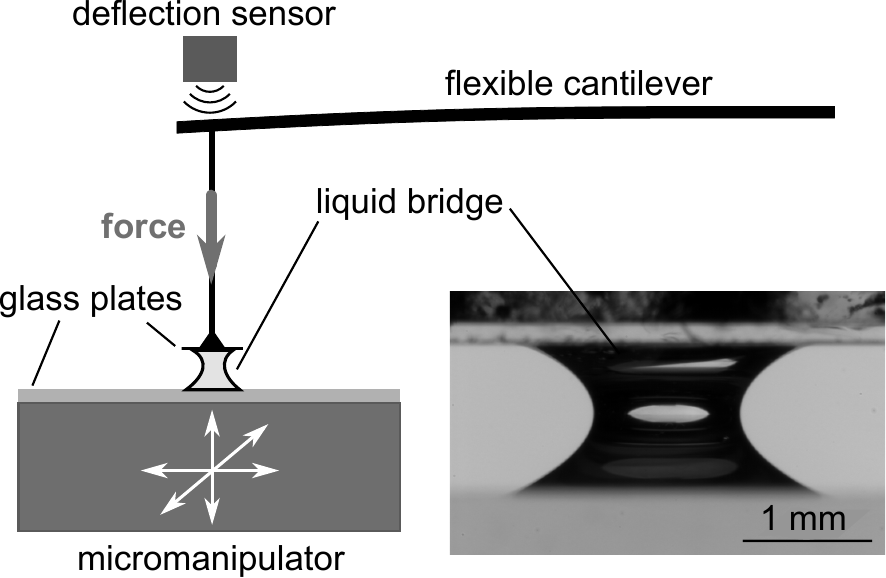}
  \caption{Drawing of the bridge tensiometer setup. Inset: Example of picture of a carbopol bridge. The white stain in the middle is a deformed image of the flat LED light situated in the back of the setup.}
  \label{fig:setup}
\end{figure}

The home made bridge tensiometer \cite{mgharbel_measuring_2009} (figure \ref{fig:setup}) allows to measure the surface tension of fluids. It consists in two horizontal glass plates, between which a droplet of the liquid of interest (a few microliters) is deposited. The liquid forms a capillary 
bridge between the two plates. The bottom plate is attached to a micromanipulator, so that its position can be 
adjusted by the operator. The force applied by the bridge on the top plate is recorded through a flexible 
cantilever equipped with an electromagnetic deflection sensor. A high resolution camera (Pixelink PL-A686M) coupled to an optical magnifier allows to take pictures of the bridge. An example of picture is shown in figure \ref{fig:setup}.

Both top and bottom plates must be perfectly cleaned to avoid line pinning which could deform the axisymmetrical bridge, and to avoid polluting the fluid with dust or surfactants. Before each series of measurements, the bottom plate is always thoroughly cleaned in a plasma cleaner. The small top plate is dipped in piranha solution (1 part of hydrogen peroxyde in 2 parts of concentrated sulfuric acid) and rinsed with deionized water.

\begin{figure}[floatfix,h]
\centering
  \includegraphics[width=6cm]{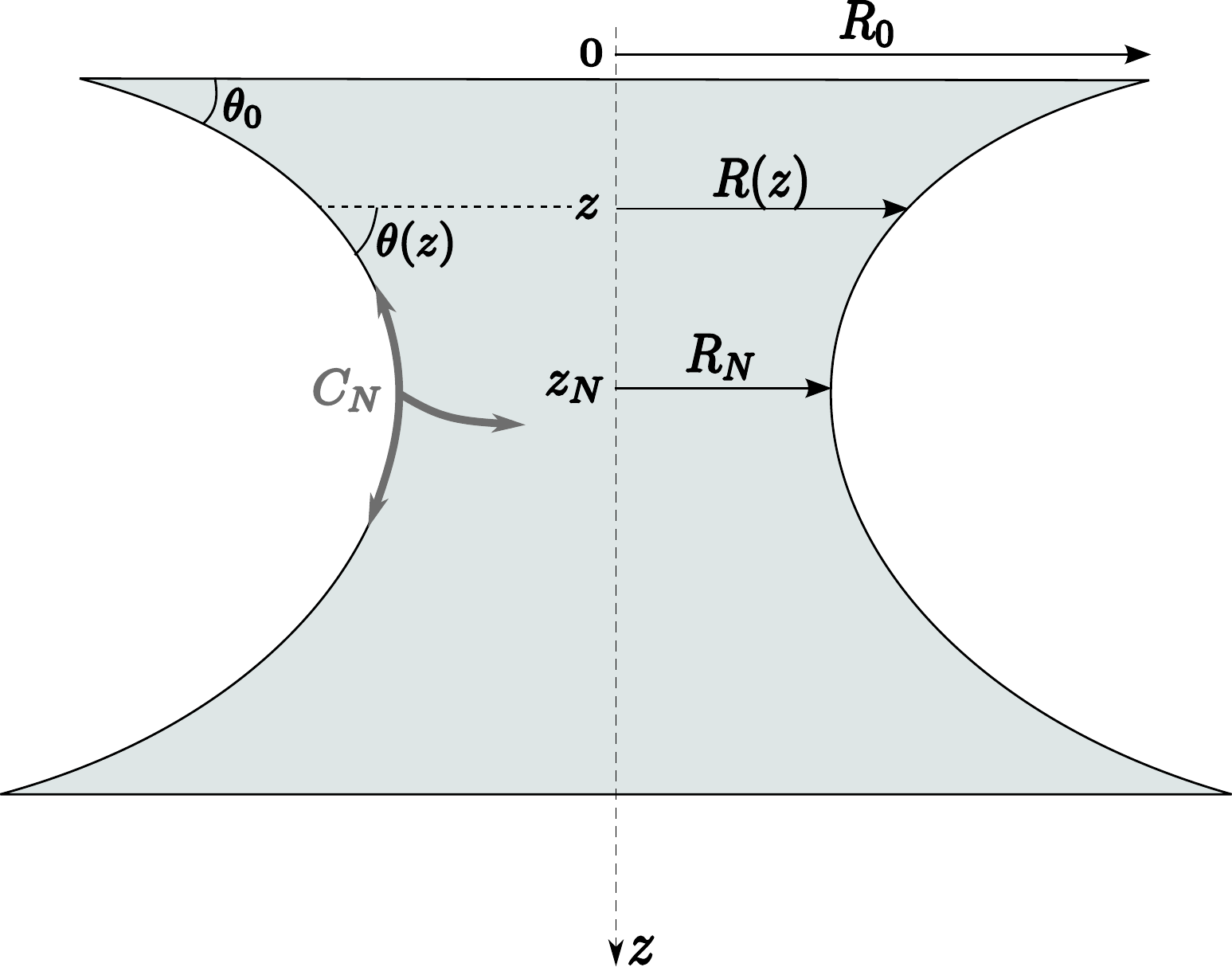}
  \caption{Definition of the main geometrical parameters.}
  \label{fig:nota}
\end{figure}

\subsubsection{Force balance.~~} At equilibrium the force on the cantilever ($F$) and the geometry of the bridge are directly linked via the surface tension of the fluid ($\gamma_{LV}$). More precisely the force measured by the cantilever is the sum of the pressure force at the liquid-plate interface and of the capillary force at the perimeter of this interface \cite{fortes_axisymmetric_1982}:
\[ F= -\pi R_0^2 \times \Delta p + 2\pi R_0 \sin\theta_0 \times \gamma_{LV} \]
where $R_0$ and $\theta_0$ are the radius and the contact angle defined on figure \ref{fig:nota}, assuming cylindrical symmetry, and $\Delta p$ is the pressure difference between the fluid and the atmosphere.

It must be noticed that the same force balance can be done at each height $z$ of the bridge, and especially at the neck ($z_N$) where $\sin\theta(z_N)=1$. However to account for gravity, it is necessary to add the weight of fluid above $z_N$ (denoted $W$) to the force balance:
\[ F= W-\pi R_N^2 \times \Delta p + 2\pi R_N \times \gamma_{LV} \]
$R_N$ being the radius of the bridge at the neck.

Finally Laplace's law allows to replace the pressure difference $\Delta p$ with $\gamma_{LV}\times C_N$, $C_N$ being the total curvature of the surface at the neck:
\begin{equation}
 F-W= \gamma_{LV} \times (2\pi R_N -\pi R_N^2 C_N)  \myeq \gamma_{LV} \times L
\label{eq:main}
\end{equation}

\subsubsection{Measurement protocol.~~} To form the bridge, a droplet is deposited on the bottom plate which is then moved upwards until contact of the liquid with the top plate. Generally the liquid spreads on the whole upper plate and the two plates are stuck together, so the bridge must always be stretched before the beginning of the measurement. During the experiment, the bridge is stretched or compressed by changing the position of the bottom plate and then let to equilibrate. Because of evaporation the force is never completely constant, but the force value and the picture are saved when the force evolution is sufficiently slow (about 1 $\mu$N per second) compared to the total force step (of the order of 100 $\mu$N in a few seconds). A typical example of force step is shown in figure \ref{fig:force} of Appendix A.

For each aspect ratio of the bridge, the geometric parameter $L$ (defined in equation \ref{eq:main}) is computed from the picture. The outline of the bridge profile is extracted from the image and fitted with a high-order (11) polynom. It is necessary to get a rather smooth profile because it is derivated twice to compute the curvature, but the polynom must nevertheless follow the real profile as faithfully as possible. The curvature of the surface is computed as:
\begin{equation*}
C(z)=\dfrac {1/R(z)}{(1+R'(z)^2)^{1/2}} - \dfrac {R''(z)}{(1+R'(z)^2)^{3/2}}
\end{equation*}

The force is obtained from the cantilever deflection after calibration. The weight $W$ is a small correction to the force and it comes from the calculated volume of fluid above $z_N$ (obtained by integration of the profile) times the fluid density. This way, $F-W$ can be plotted as a function of the corresponding parameter $L$. In the case of a simple fluid, a proportional relation is expected (see relation \ref{eq:main}), the slope being the surface tension value $\gamma_{LV}$.

\section{Experimental results}
\label{sec:exp}
\subsection{Simple fluids}
In order to validate the setup, the experiment was first performed with simple fluids. The protocol was always the same, testing stretching as well as compression to check the influence of the dynamics history on the results.

\begin{figure}[floatfix,h]
\centering
  \includegraphics[width=0.9\columnwidth]{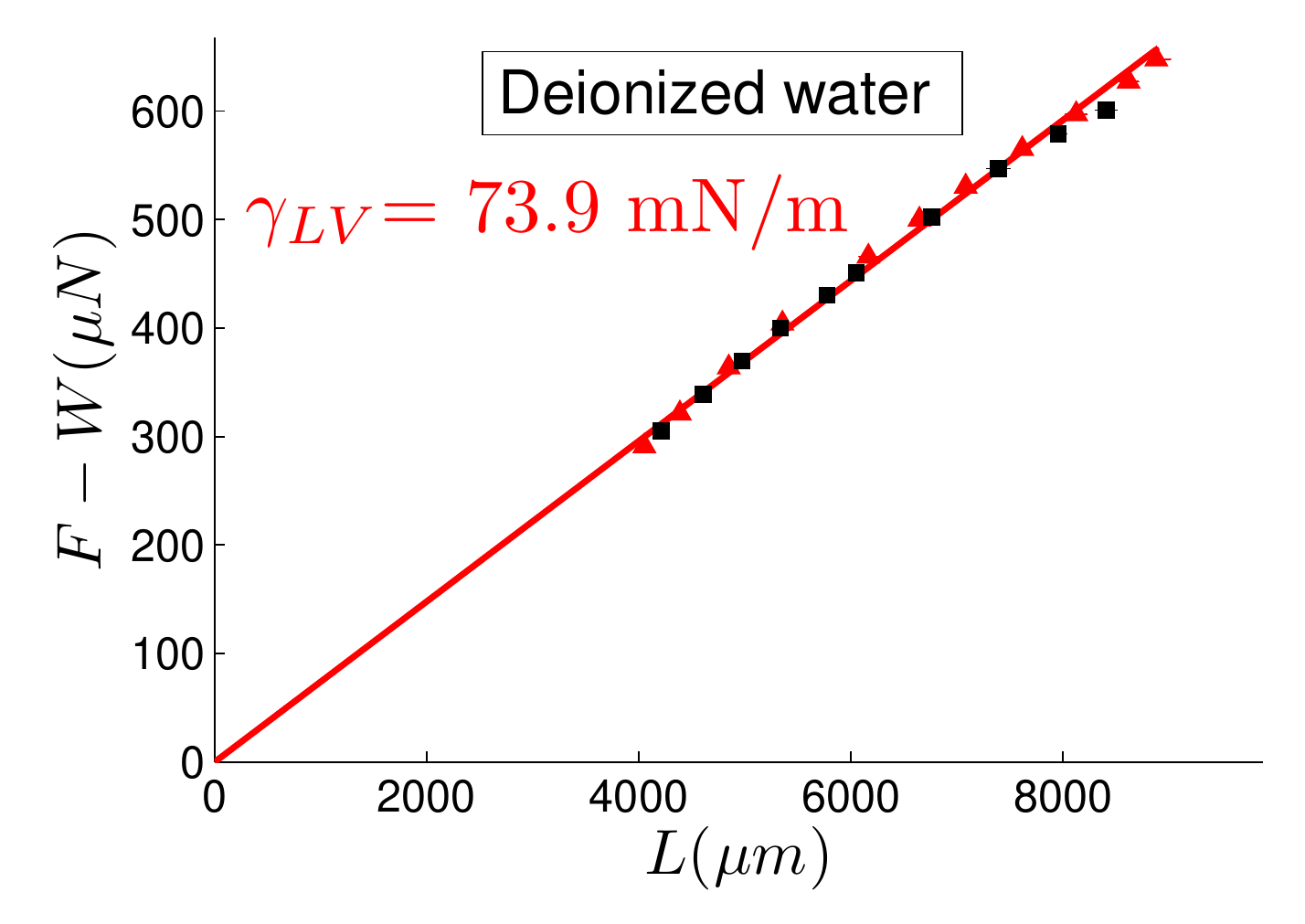}\\\includegraphics[width=0.9\columnwidth]{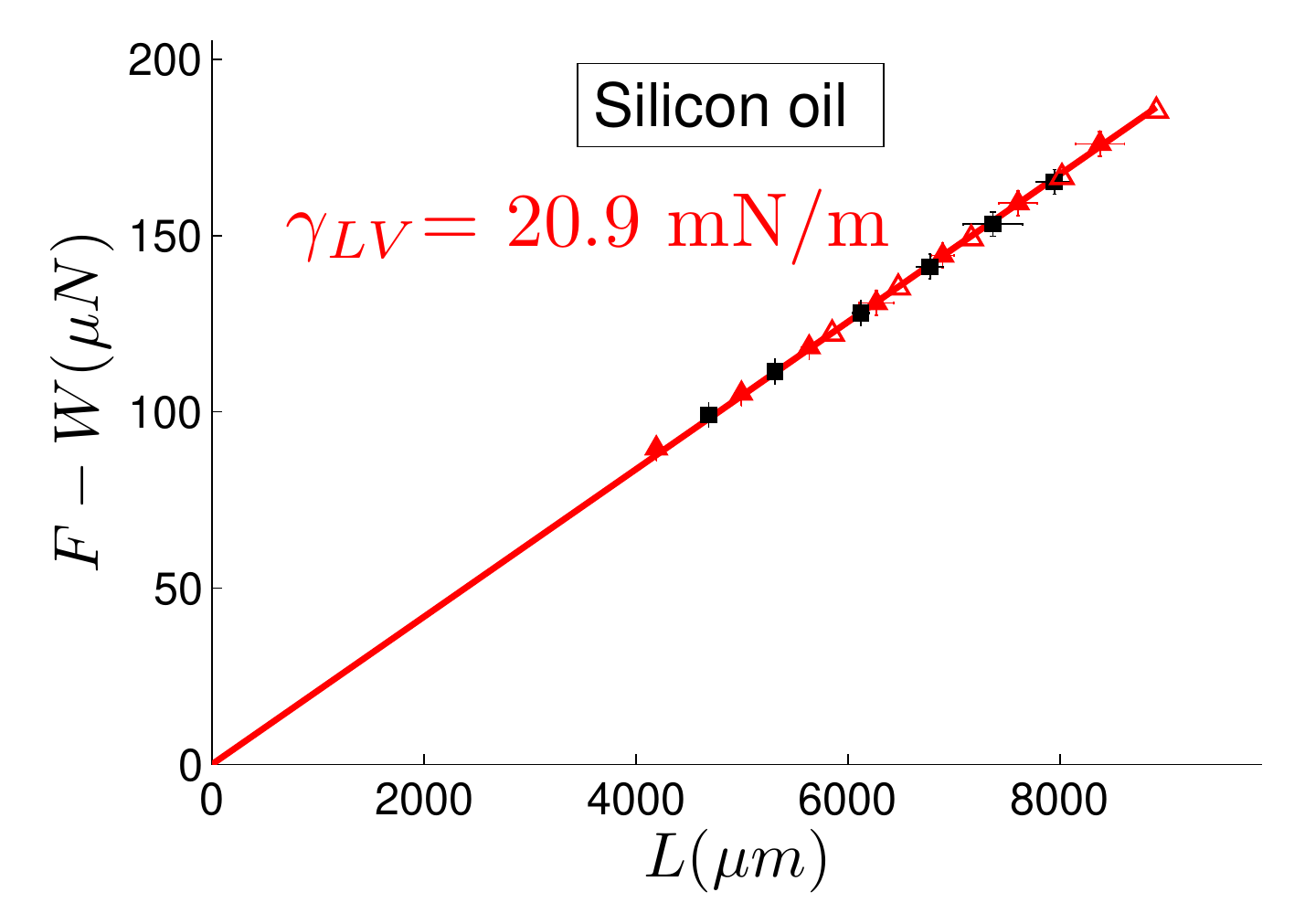}
  \caption{Force-L plot for deionized water and silicon oil. The force $F-W$ is proportional to $L$ and the slope corresponds to the surface tension of the liquid. Red triangles and black squares respectively stand for stretching and compression steps.}
  \label{fig:simple}
\end{figure}

With pure water and silicon oil, the force-$L$ plot indeed shows a proportional relation (see figure \ref{fig:simple}) and the slopes correspond to respective surface tensions of $(74 \pm 1)$~mN/m and $(21 \pm 1)$~mN/m. The expected surface tensions are 73.0~mN/m and 21.0~mN/m (at 18$\degC$). The agreement is very good, with precision comparable to usual surface tension measurement methods \cite{zuidema_ring_1941,heertjes_determination_1971,girault_drop_1982,christian_inverted_1998}.

Great care must be taken to avoid hysteresis of the contact line. A treatment was applied on the glass plates \cite{krumpfer_contact_2010} to minimize hysteresis before the measurements with oil and water.

\subsection{Carbopol}
Carbopol is a yield stress fluid, i.e. it cannot flow if the applied stress is below the yield stress ($\sigma_Y$). In particular it is necessary to make small droplets such as the surface tension induced pressure dominates over the yield stress. This can be quantified by a dimensionless number comparing the yield stress energy $\sigma_Y\times r^3$ ($r$ being the characteristic size of the system) and the surface energy $\gamma_{LV}\times r^2$. For example, in a droplet of gel with $\sigma_Y=20$~Pa and $\gamma_{LV}=60$~mN/m, the excess pressure in the liquid due to capillary forces must be greater than 20~Pa for the droplet to flow and adopt a spherical shape. Therefore we expect that an isolated droplet of this fluid of radius smaller than 6~mm ($r<2\gamma_{LV}/\sigma_Y$) is spherical. More generally if the curvature of the surface is greater than $\sigma_Y/\gamma_{LV}$, the shape of the liquid surface should be controlled by surface tension. This is the reason why only small capillary bridges (radius of the order of the millimeter) were studied in this experiment.

\begin{figure}[floatfix,h]
\centering
  \includegraphics[width=0.85\columnwidth]{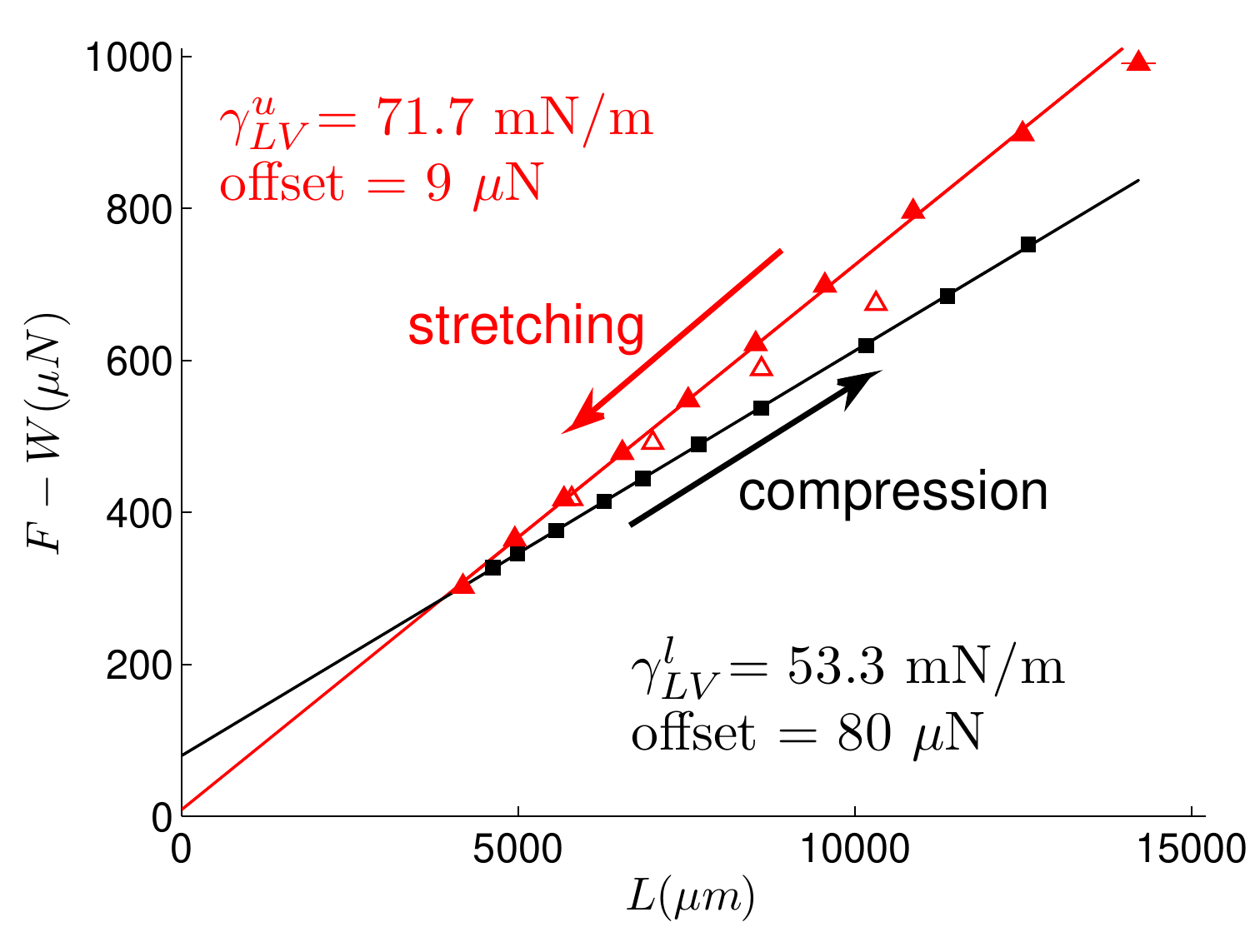} \put(-170,112){a)}\\
\includegraphics[width=0.85\columnwidth]{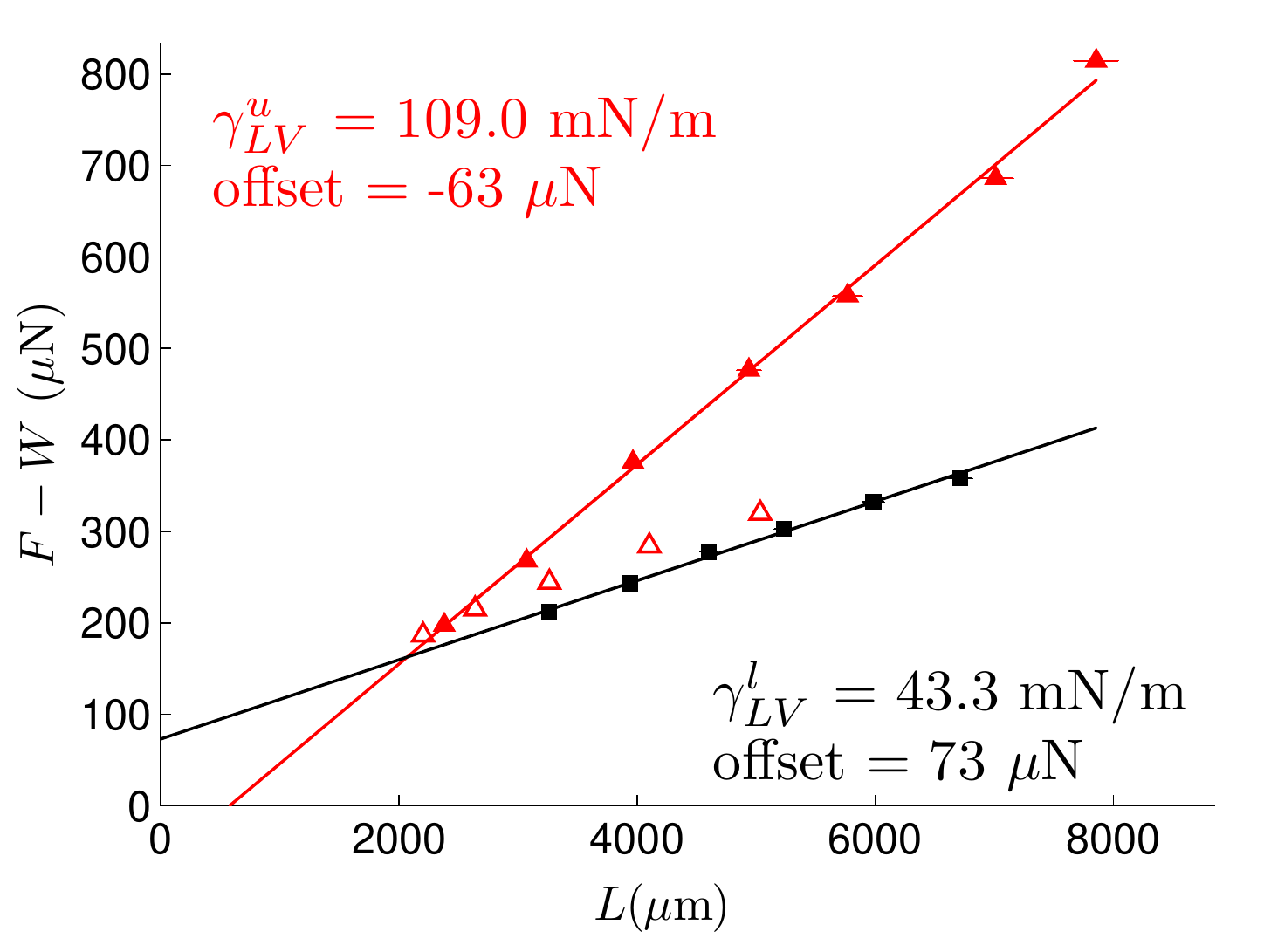} \put(-175,110){b)}
  \caption{Force-L plot for two different carbopol samples: a) 0.25\% (HS), yield stress $\sigma_Y=5$ Pa ; b) 1\% (MS), yield stress $\sigma_Y=19$ Pa. The solid (resp. empty) red triangles stand for the first (resp. second) series of stretchings, the black squares for the series of compressions. The linear fit values are written in the figure.}
  \label{fig:Vexample}
\end{figure}

As for the simple fluids we started with a series of stretchings and then a series of compressions. Most of the time these were followed by a second series of stretchings. It must be clear that compressed bridges correspond to large values of $L$ and that a stretching is transposed into shift to the left side of the force-$L$ plot. On the contrary a shift to the right of the plot is a compression (see figure \ref{fig:Vexample}).

Typical force-$L$ plots for carbopol 0.25\% (HS, $\sigma_Y=5$ Pa) and carbopol 1\% (MS, $\sigma_Y=19$ Pa) are reproduced in figure \ref{fig:Vexample}. We observe that the points do not all align on a single line. The solid red triangles correspond to the first series of stretching, starting at the top-right angle of the plot. The red line is the linear fit of these points, and its slope is denoted $\gamma_{LV}^{\, u}$. The black squares correspond to the series of compressions. They align on a second line, whose slope $\gamma_{LV}^{\, l}$ is always smaller than for the stretched points.  This behavior is reproducible for every sample of carbopol, and the greater the yield stress, the wider the difference of slopes between the two sets of points.

\begin{figure}[floatfix,h]
\centering
 \includegraphics[width=0.9\columnwidth]{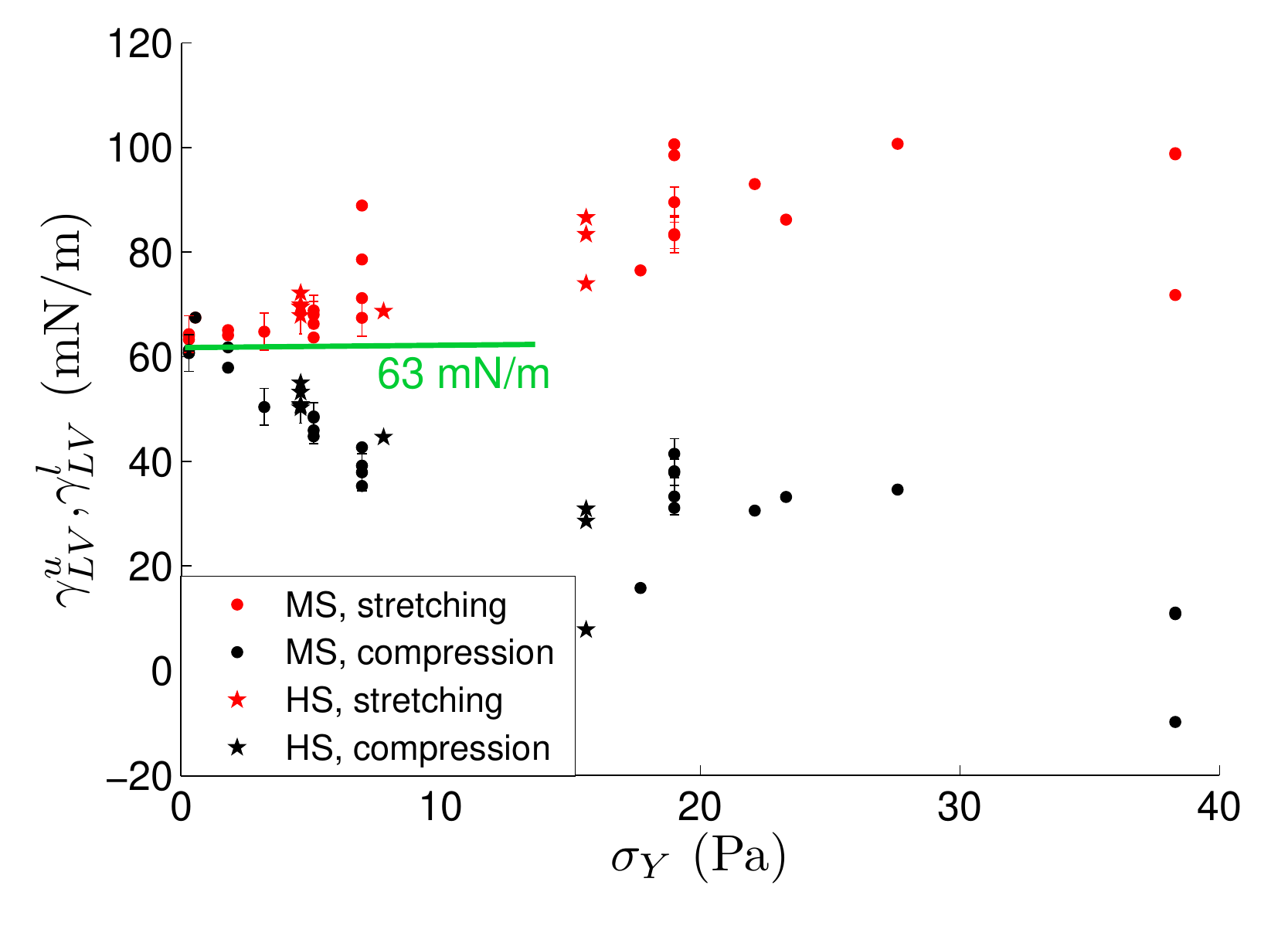}
  \caption{Upper (red) and lower (black) slopes of the force-$L$ plots plotted as a function of the yield stress. The green line is a guide for the eyes, indicating the mean surface tension of vanishing yield stress carbopols. The error bars indicate the averaged points (see text). The error on all the other points is $\pm 5$ mN/m.}
  \label{fig:gam_vs_ys}
\end{figure}

\begin{figure}[floatfix,h]
\centering
 \includegraphics[width=0.9\columnwidth]{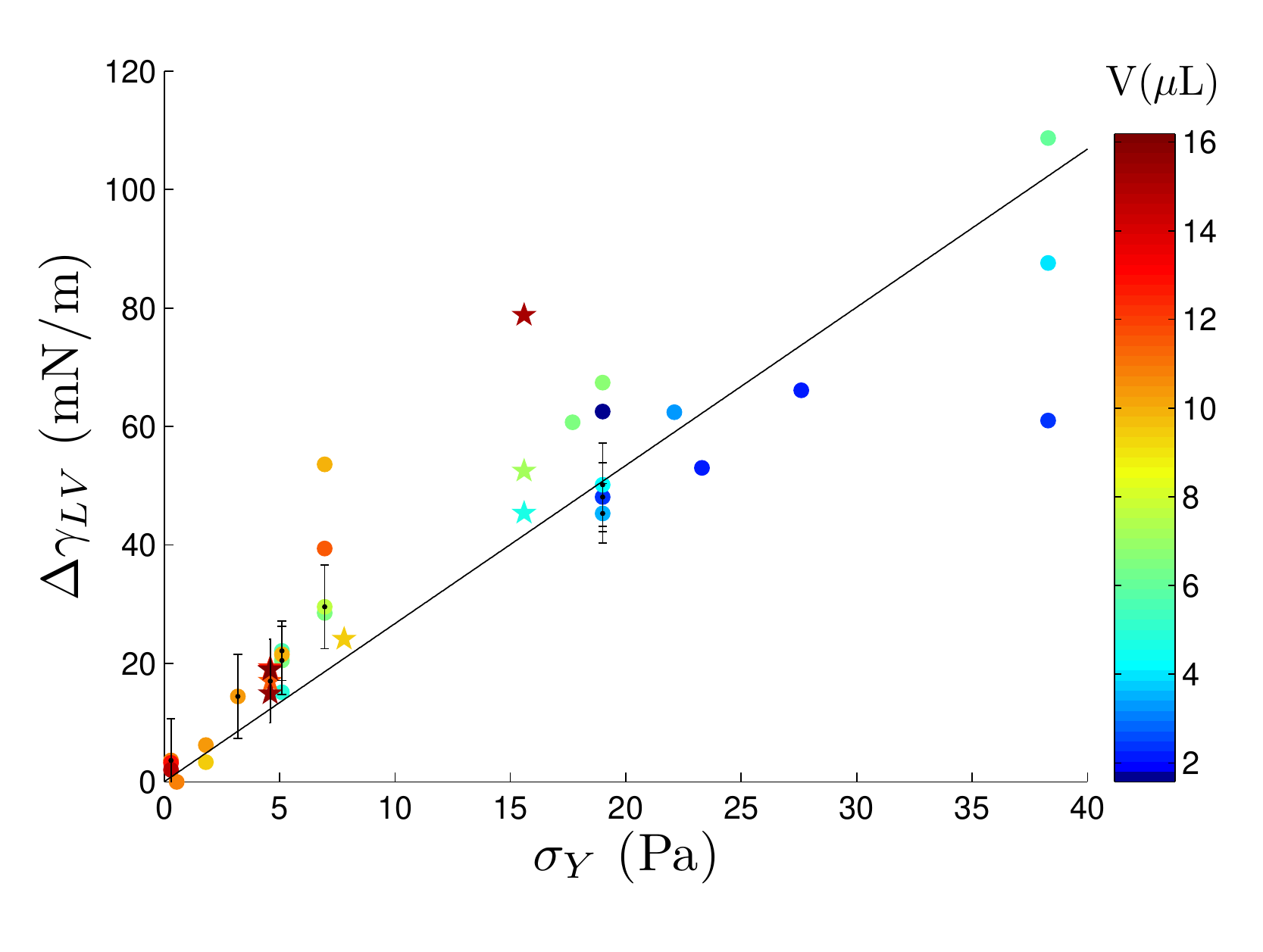}
  \caption{Difference $\Delta\gamma_{LV}=\gamma_{LV}^{\,u} -  \gamma_{LV}^{\,l}$ of the force-$L$ plots slopes, as a function of the yield stress of the samples. Each point color represents the volume of the droplet. Star-shaped points stand for HS carbopol and dots for MS carbopols. The line indicates a linear fit whose correlation coefficient $\mathrm(R)^2$ is only 0.74.}
  \label{fig:V_vs_ys}
\end{figure}

To confirm the influence of the yield stress on the apparent surface tension, we performed several experiments, varying $\sigma_Y$ between 0.3~Pa and 38~Pa. This could be achieved by varying either the polymer concentration or the stirring. Hand-stirred carbopols have indeed a much greater yield stress than machine-stirred carbopols of same concentration. 

For a few samples the experiment was performed with several droplet volumes between 2~$\mu$L and 15~$\mu$L. Moreover for two of them, 10 identical measurements were carried out in order to evaluate the dispersion of the effective surface tension values. The standard deviation of the results is of about 5~mN/m for given yield stress and volume.

Figure \ref{fig:gam_vs_ys} shows the values of the upper and lower slopes as a function of the sample yield stress, and each point is an average on 1 to 4 droplets of similar volume (within 1 $\mu$L steps) and yield stress (within 1 Pa steps). It can be observed that the upper slope increases with the yield stress while the lower slope decreases. For vanishing yield stress, they both converge to 63~mN/m.

Figure \ref{fig:V_vs_ys} is a plot of the slopes difference $\Delta\gamma_{LV}=\gamma_{LV}^{\,u} -  \gamma_{LV}^{\,l}$ vs $\sigma_Y$, with the same average as before, and the droplet volume is represented by the point color. It confirms the monotonic dependence of the slopes difference with the yield stress, and it also shows that greater $\Delta\gamma_{LV}$ often correspond to larger drops, for a given yield stress.

In both figures the star-shaped points stand for HS carbopol samples and the other points for MS samples. The averaged points are indicated by error bars.

\begin{figure}[floatfix,h]
\centering
  \includegraphics[width=0.85\columnwidth]{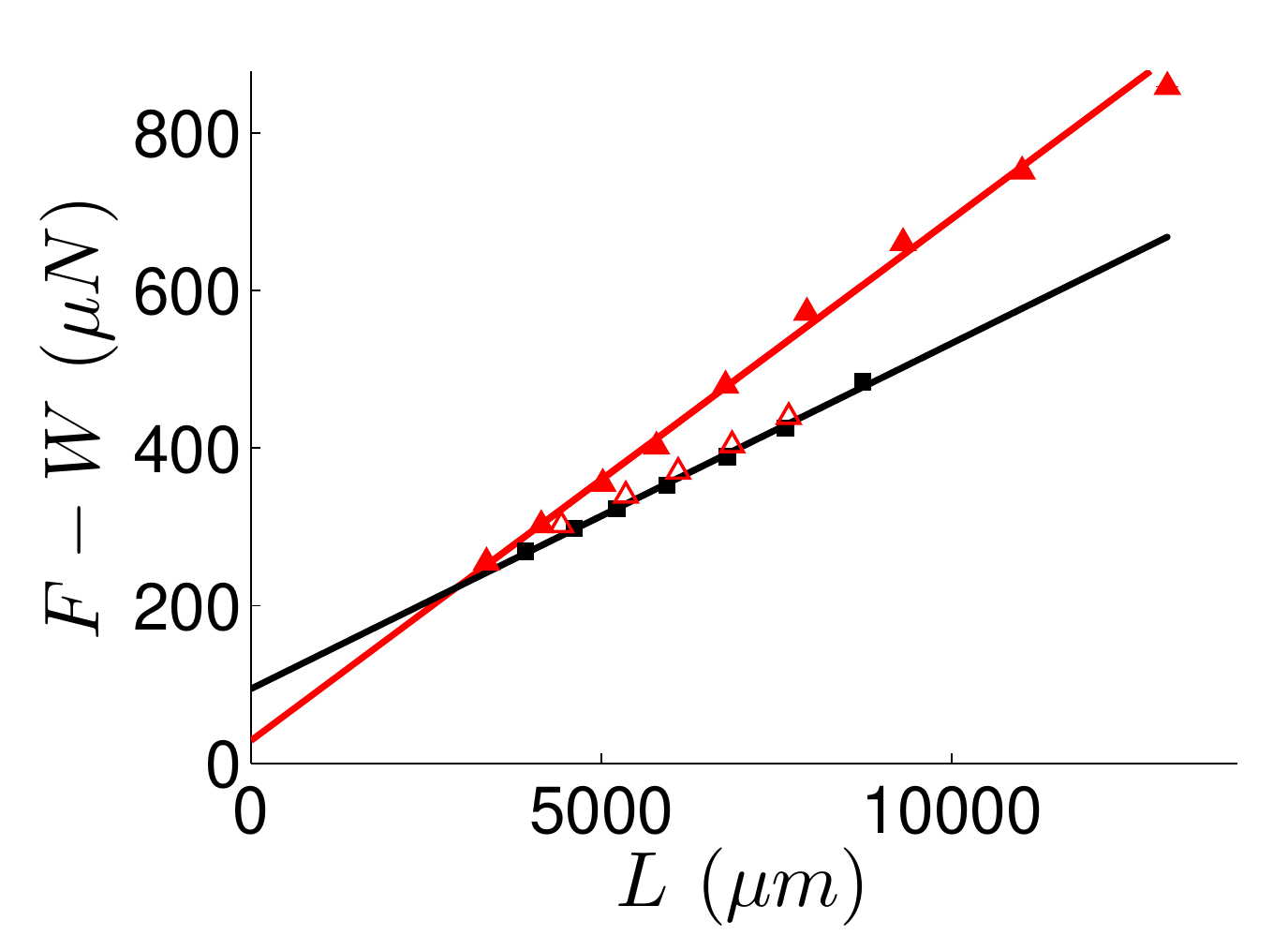} \put(-160,130){a)}\\ \includegraphics[width=0.85\columnwidth]{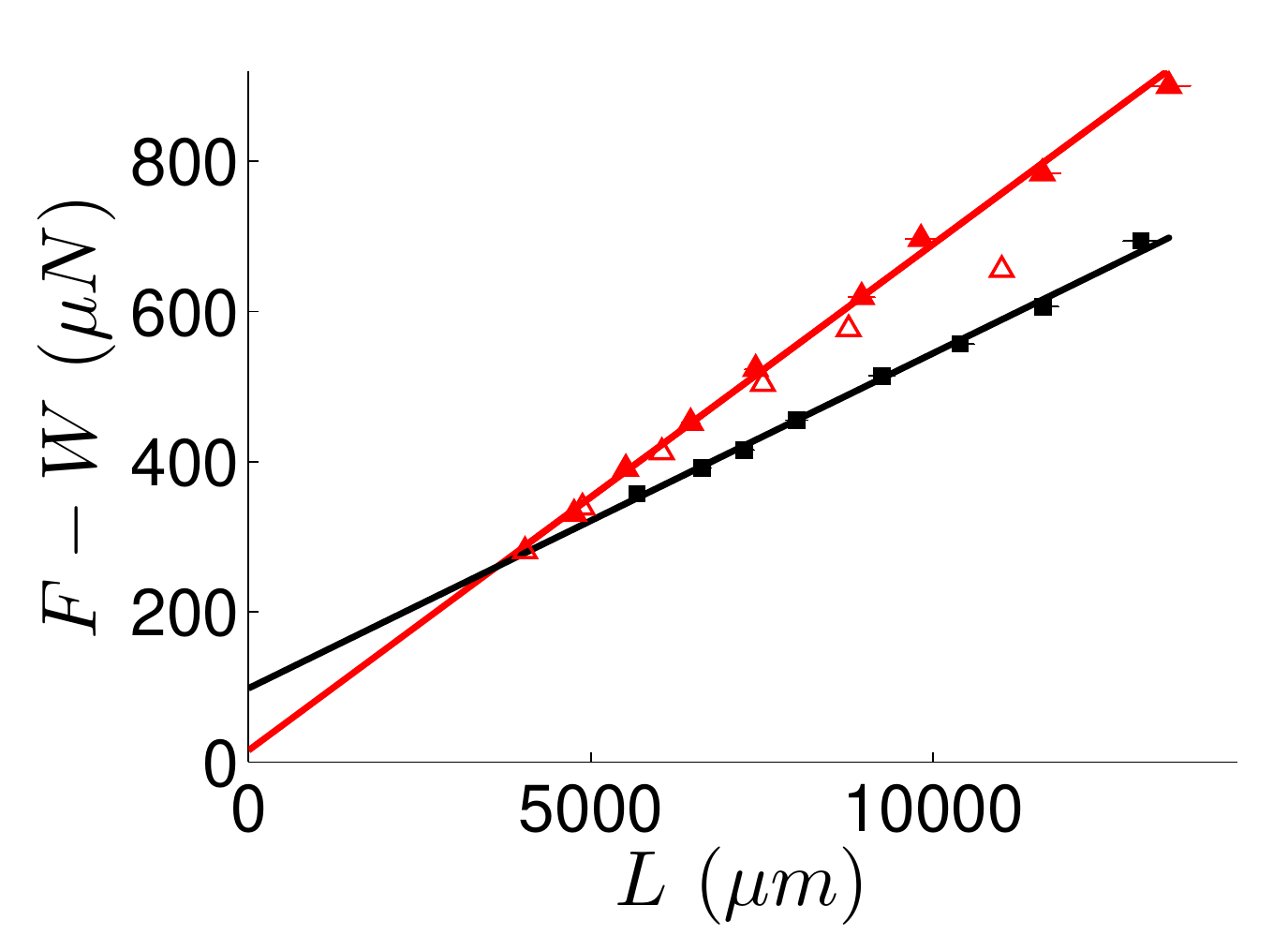}\put(-160,130){b)}\\
  \caption{Example of two force-$L$ plots for carbopols of same yield stress $\sigma_Y=7$~Pa and different elastic moduli. a) MS carbopol, $G'=20$~Pa. b) HS carbopol, $G'=45$~Pa.}
  \label{fig:elasticexp}
\end{figure}

Finally, as shown on figure \ref{fig:elasticexp}, the shape of the second stretching cycle (empty red triangles) varies from one experiment to another. The second stretching set of points joins the first stretching line (red) faster when the elastic modulus of the carbopol is higher, for equal yield stresses.

\section{Elastoplastic model}
To understand the influence of the different parameters in our experiment, we have developed a simple model. The goal is to understand the role of the flow history on the curves obtained with a yield-stress fluid.

Because the experiments clearly show an influence of both the yield stress and the elasticity of the fluid, we consider an elastoplastic fluid: below $\sigma_Y$ it behaves as an elastic solid, and at $\sigma_Y$ it flows until it reaches a stationary state. We neglect the consistency $K$ of the Herschel-Bulkley model as the time evolution of the force is not investigated here, only the final state. The model mimics the experimental protocol and explores the influence of the elastic deformation on the stress state of the bridge for either stretching or compression and different initial conditions. To be able to calculate the stress, we consider two limiting simplified geometries, the filament and the pancake.

This model allows to faithfully reproduce the experimental results and thus to explain the observations exposed in part \ref{sec:exp} as explained below.

A drop of viscoplastic liquid with yield stress $\sigma_Y$ and shear elastic modulus $G'$ is considered. The drop has a nearly cylindrical shape with height $h$ and neck radius $R_N$, so that the volume of the drop is $V \approx \pi R_N^2 h$. We denote $\theta_0 = 30 ^\circ$ the contact angle, which is roughly the contact angle observed in the experiments. The total curvature is assumed to be constant along $z$, and the geometric parameter $L$ is approximated by
\begin{align*}
L &= 2 \pi R_N - \pi R_N^2 \left ( \frac{1}{R_N} - \frac{2 \cos \theta_0}{h} \right )\\ &= \pi R_N + \frac{2 \pi R_N^2 \cos \theta_0}{h}\\  &\approx \pi  \sqrt{\frac{V}{\pi h}} + \frac{2 V \cos \theta_0}{h^2}
\end{align*}

For a given volume $V$, the filament (resp. pancake) geometry corresponds to heights $h \gg (V/\pi)^{1/3}$ (resp. $h \ll (V/\pi)^{1/3}$). The volume is fixed to $V = 10$~mm$^3$, as often encountered in experiments. This corresponds to $(V/\pi)^{1/3} \approx 1.5$~mm. As this is the typical experimental value of $h$, the experiments do not correspond to any of these limiting geometries (filament or pancake), but to an intermediate regime where $h\sim R$. But as discussed later, we show that the results of the model do not qualitatively depend on the chosen geometry. This is the reason why we choose to explain the model in details in the filament geometry only. The pancake geometry calculations are nevertheless presented in Appendix B.

\subsection{Filament geometry}
\label{par:model}
In this geometry, usually encountered in capillary thinning or filament-stretching devices \cite{anna_elasto-capillary_2001,mckinley_filament-stretching_2002}, elongational deformation and normal stress (and not shear) are assumed to be dominant. The characteristic stress is the normal stress  $\sigma = \sigma_{zz} - \sigma_{rr}$.
We look at small height variations $\Delta h$. The corresponding step in deformation is:
 $$ \Delta \varepsilon =  \frac{\Delta h}{h} $$ and the stress before each step is denoted $\sigma_0$.

In the elastoplastic hypothesis, and taking into account the tensorial formulation of the stress tensor \cite{coussot_saffman-taylor_1999}, the new stress after a step is given by the following function: 
$$
\sigma  = \Bigg\{
  \begin{tabular}{lcl}
   $-\sqrt{3} \sigma_Y$ & \ & if $ \sigma_0 + 3 G'  \Delta \varepsilon < - \sqrt{3} \sigma_Y $ \\
  $\sigma_0 + 3 G' \Delta \varepsilon$ & \ & if $-\sqrt{3} \sigma_Y  < \sigma_0 + 3 G'  \Delta \varepsilon < \sqrt{3} \sigma_Y $ \\
   $+\sqrt{3}\sigma_Y$ & \ & if $ \sigma_0 + 3 G'  \Delta \varepsilon > + \sqrt{3} \sigma_Y $ \\
  \end{tabular}
$$

 Finally, the normal elastoplastic force applied on the cantilever is evaluated at each step:
 $$F_\text{ep} = \sigma \pi R_N^2 = \sigma \frac{V}{h}$$ 

In the experiments the drop is initially stretched so the initial stress is set to $+\sqrt{3} \sigma_Y$. Then successive steps of deformation $\Delta h = 0.3$~mm are applied to the model drop, starting with stretching from $h=1.5$~mm to $h=4.5$~mm, then compressing and finally stretching again.

For each step, the total traction force, which is the sum of the capillary force $\gamma_{LV}L$ and the elastoplastic one $F_\text{ep}$, is calculated for $\gamma_{LV} = 60$~mN/m and $\sigma_Y=2$~Pa or 5~Pa, using the following approximation of $L =\pi  \sqrt{V/(\pi h)}$. Several values of the elastic modulus $G'$ are tested: $G'/ \sigma_Y = 0.5$, 2 and 8.

\begin{figure}[floatfix,h]
\centering
\includegraphics[width=0.9\columnwidth]{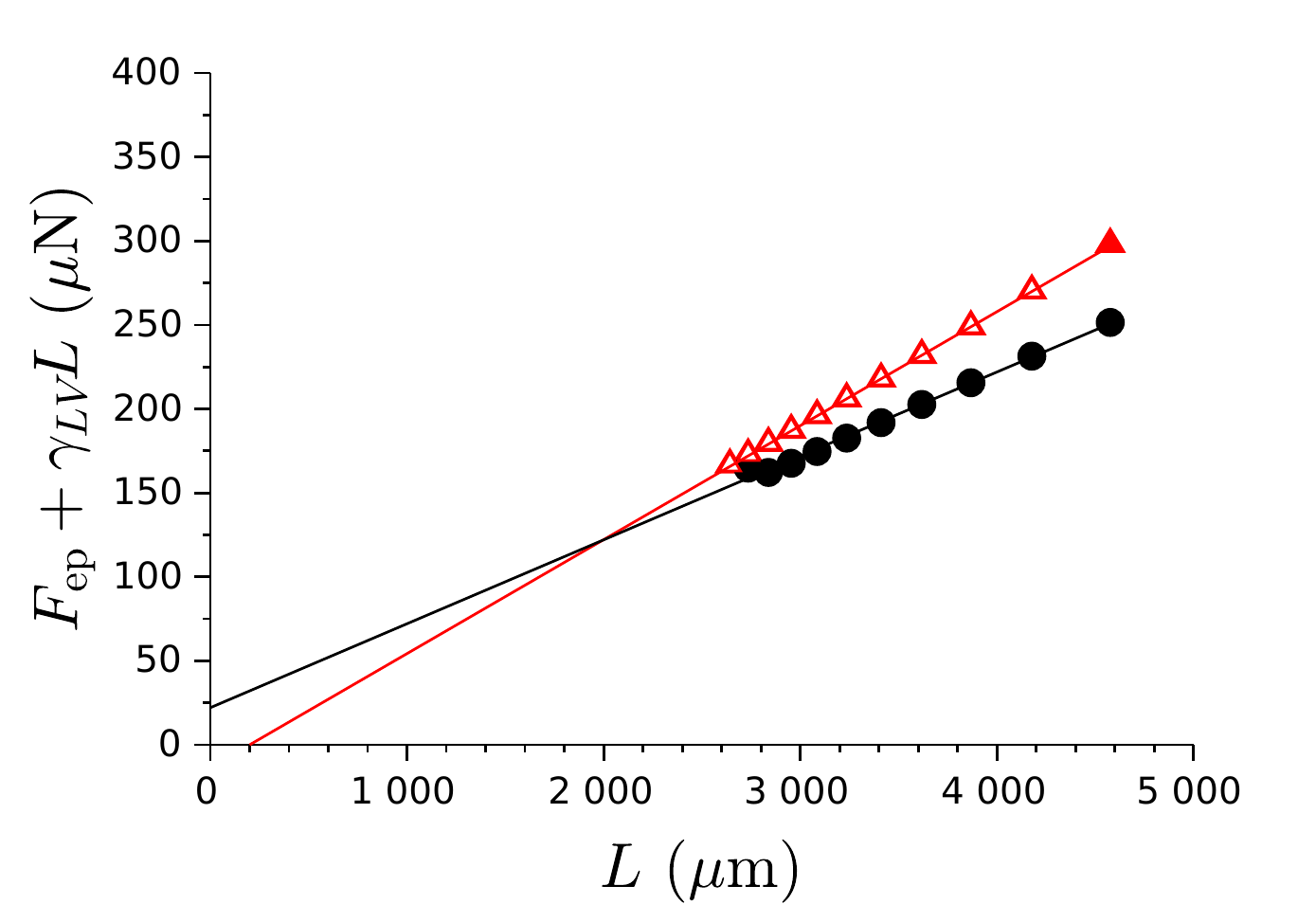}\put(-175,130){a)}\\
\includegraphics[width=0.9\columnwidth]{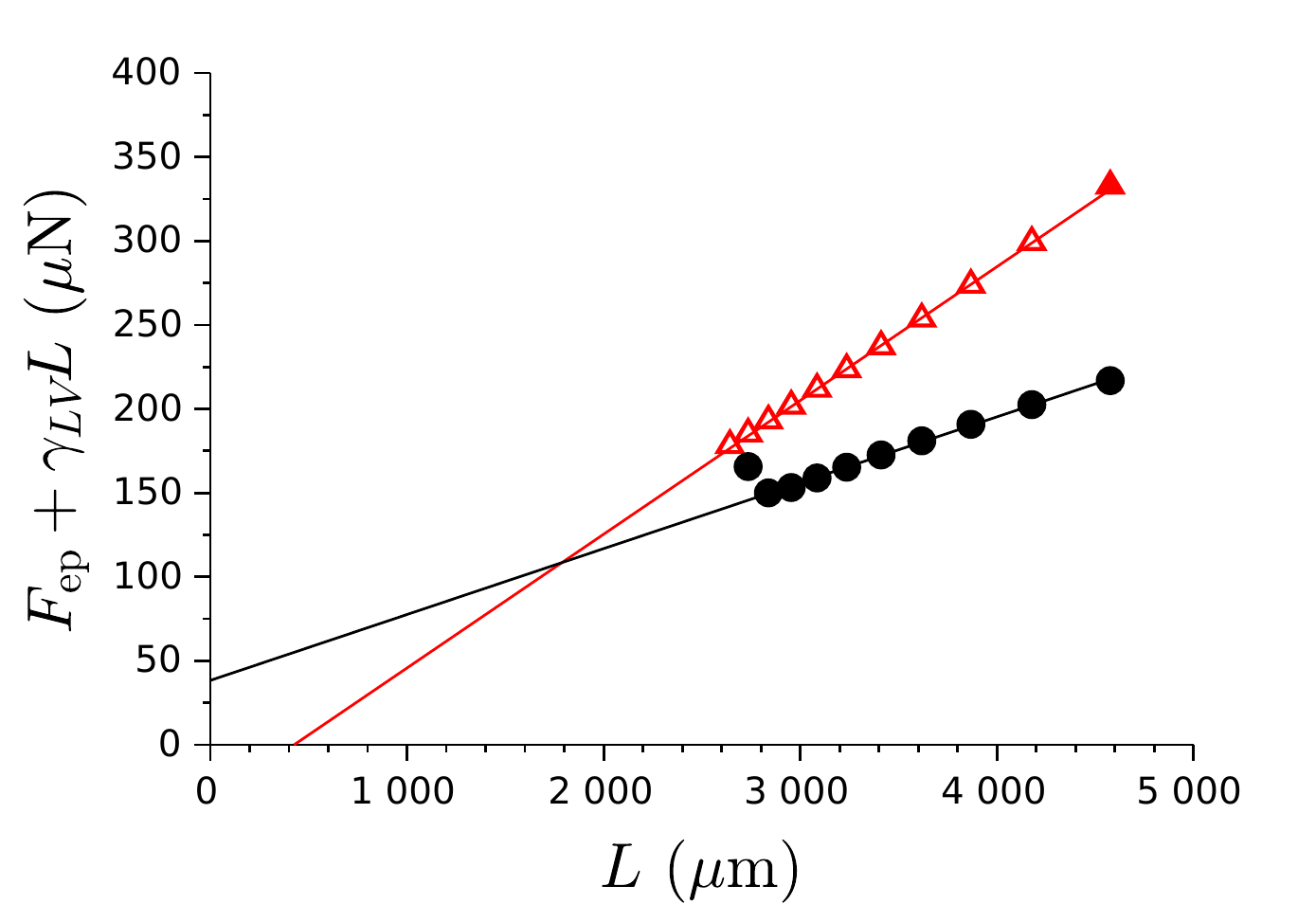}\put(-175,130){b)}\\
\caption{Results from the model: $F$ as a function of $L$ for a filament geometry, with $G' / \sigma_Y=8$. a) $\sigma_Y=2$~Pa. b) $\sigma_Y=5$~Pa.}
\label{fig:filamentsigma}
\end{figure}

\begin{figure}[floatfix,h]
\centering
\includegraphics[width=0.9\columnwidth]{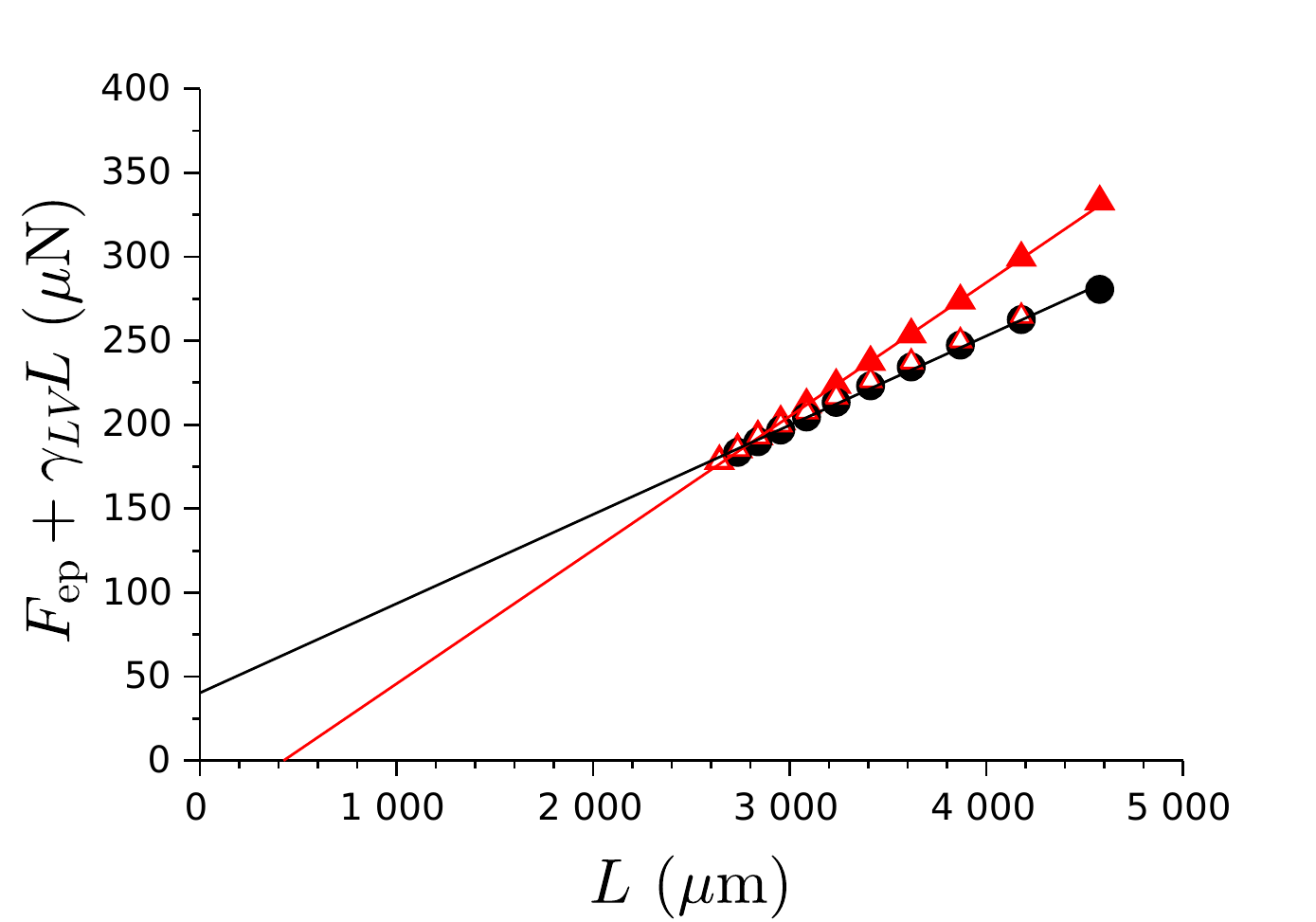}\put(-175,130){a)}\\ \includegraphics[width=0.9\columnwidth]{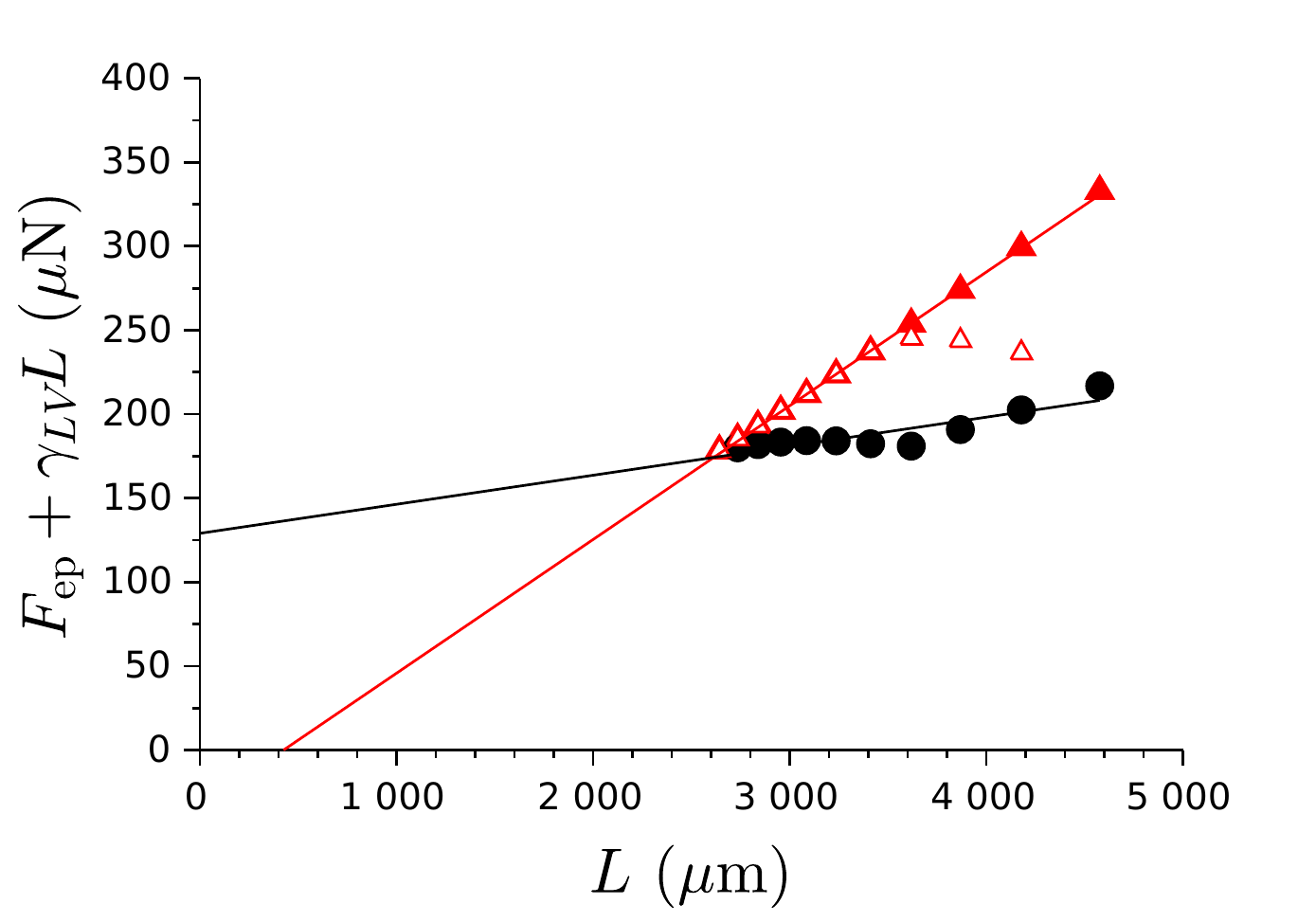}\put(-175,130){b)}\\
\includegraphics[width=0.9\columnwidth]{fig/filament_sigmaY5Pa_GsursigmaY8.pdf}\put(-175,130){c)}
\caption{Results from the model: $F$ as a function of $L$ for a filament geometry. $\sigma_Y=5$~Pa and  $G' / \sigma_Y=0.5$, 2 and 8 from a) to c).}
\label{fig:filamentGp}
\end{figure}

On figure \ref{fig:filamentsigma}, plots a) and b) differ only by the yield stress value. It is clear that the slopes difference between the two branches increases with the yield stress $\sigma_Y$. Here the $L$-range is small compared to the experiments because elongated shapes correspond to low values of $L$.

Figure \ref{fig:filamentGp} represents three typical force-$L$ plots from the model, for a given yield stress ($\sigma_Y=5$~Pa) and different elastic moduli $G'$. It shows that the elastic modulus has a strong influence on the shape of the stretching-compression cycle. First the shape of the lower branch (black symbols) changes when $G'/\sigma_Y$ increases, causing a variation of the y-intercept and of the slope of the black linear fit. Then the second stretching branch (empty triangles), as in the experiments, joins the first stretching branch (solid triangles) all the faster as $G'$ is large. The first stretching branch does not change because the normal stress is $+\sqrt{3} \sigma_Y$ all along and $G'$ plays no role here.

Note that for values of $G'/\sigma_Y$ of the order of 10 or more, the maximal elastoplastic stress ($\sqrt{3}\sigma_Y$) is reached immediately after the direction change. This means that for $G'/\sigma_Y\ge10$ the points fall on two limiting curves determined only by the yield stress. These two curves are symmetrical with respect to $F-W=\gamma_{LV}L$. This allows to find the true value of $\gamma_{LV}$ by taking the mean of the two limiting slopes.

\subsection{Pancake geometry}
We also checked the other limit of a flattened drop. In this case, the deformations and dissipation are dominantly due to shear along the $z$ direction. Therefore we cannot use a homogeneous description but we need to describe the stress profile at the wall.

The details of the calculation in the lubrication approximation can be found in the appendix B.

\begin{figure}[floatfix,h]
\centering
\includegraphics[width=0.9\columnwidth]{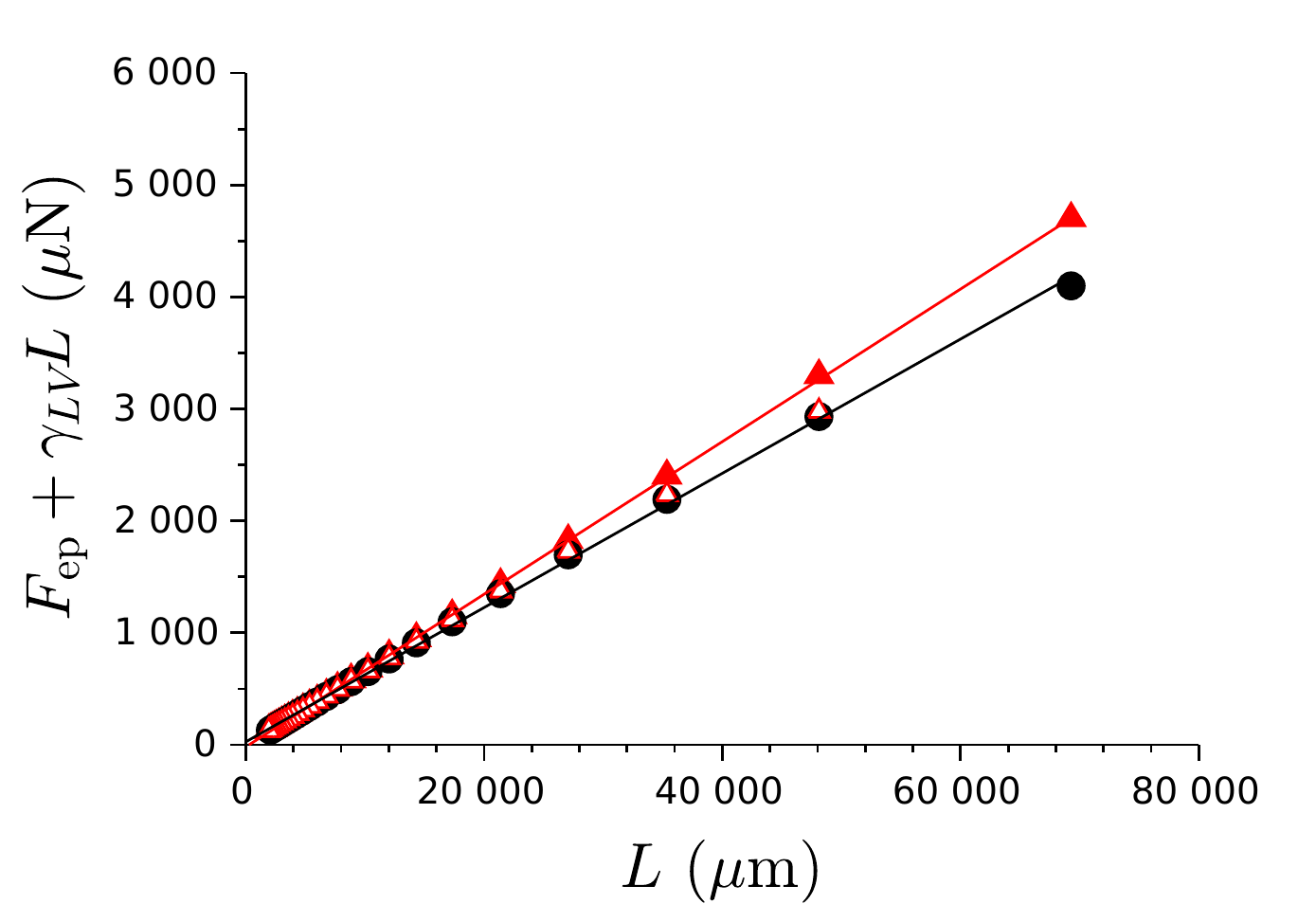}\put(-170,130){a)}\\
\includegraphics[width=0.9\columnwidth]{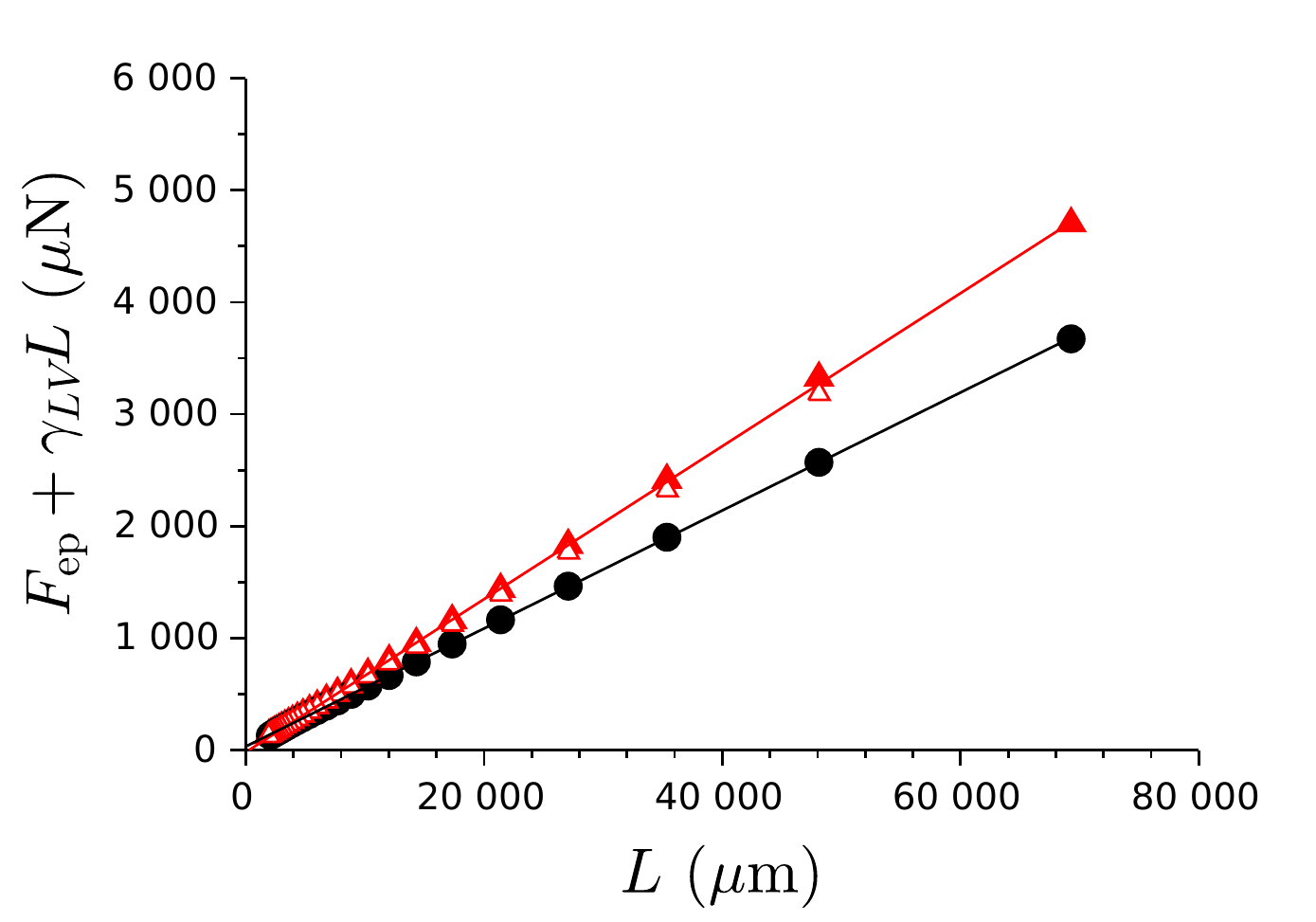}\put(-170,130){b)}
\caption{Results from the model: $F$ as a function of $L$ for a pancake geometry and $\sigma_Y= 10$~Pa. a): $G' / \sigma_Y = 0.2$.  b): $G' / \sigma_Y = 2$.}
\label{fig:pancake}
\end{figure}

What is observed with this geometry does not differ qualitatively from the case of the filament (see figure \ref{fig:pancake}): stretching (resp. compression) of the capillary bridge is associated with an increase (resp. decrease) of the apparent surface tension. The $L$-range increases, which is consistent with the displacement towards the right of the plot with compression.

\section{Discussion}
\subsection{Influence of yield stress and volume}
The model confirms the influence of yield stress on the difference of effective surface tensions. For large enough elastic moduli $G'\gg\sigma_Y$ the excess force due to the yield stress can be approximated by $\frac{\Delta\gamma_{LV}}2\times L$. In the filament geometry, assuming that the stress has reached its saturation value, this excess force can be estimated by $\sqrt3\sigma_Y\times \pi R_N^2$, and $L\approx \pi R_N$ so the slopes difference reduces to $\Delta\gamma_{LV}\propto R_N \sigma_Y$. In the pancake geometry, the excess force is about $\frac{2\pi}3 \sigma_Y \times R_N^3/h$ \cite{engmann_squeeze_2005} and $L\approx 2\pi R_N^2\cos\theta_0/h$ so the relation $\Delta\gamma_{LV}\propto R_N \sigma_Y$ still holds.

\begin{figure}[floatfix,h]
\centering
\includegraphics[width=\columnwidth]{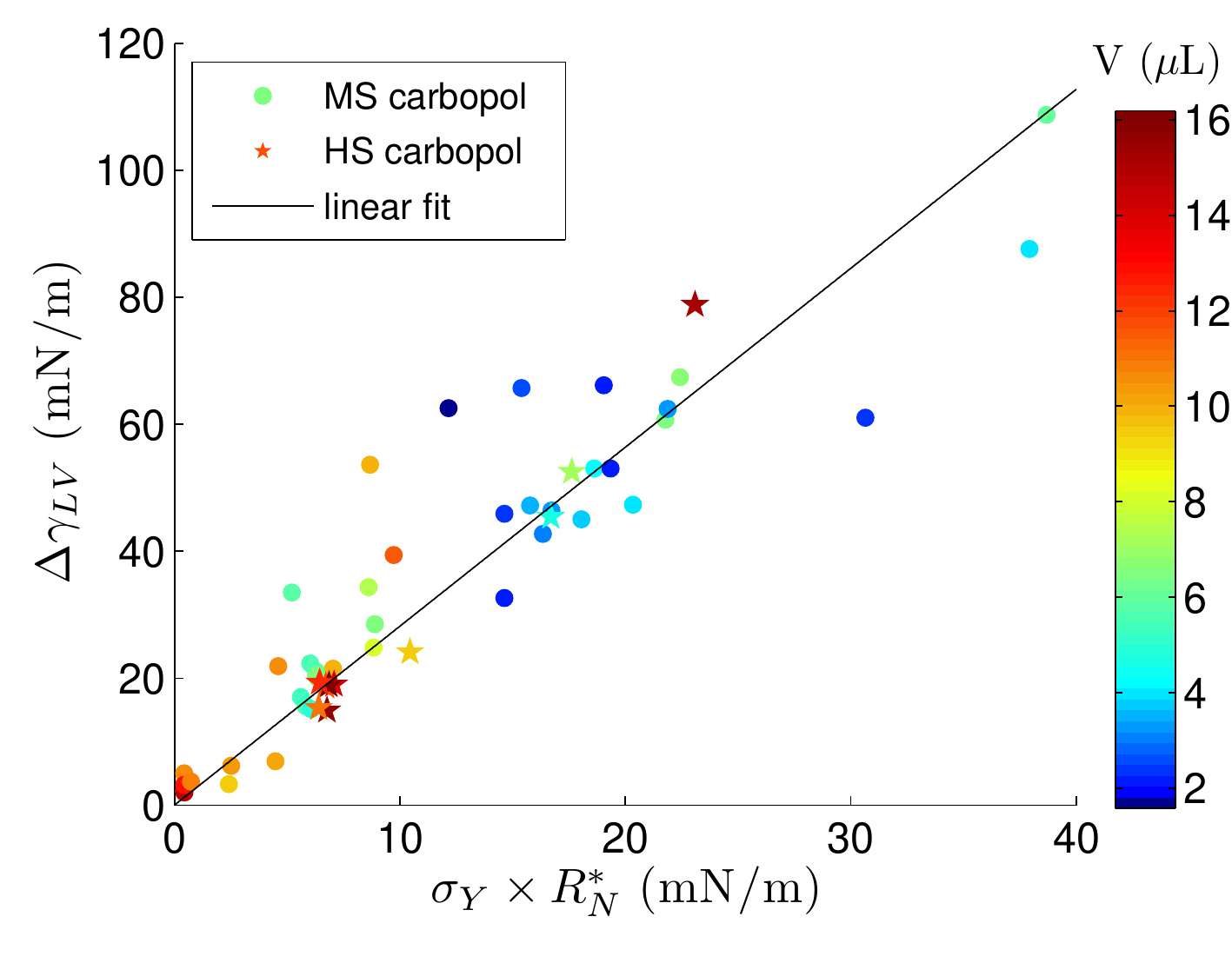}
\caption{Effective surface tensions difference $\Delta\gamma_{LV}$ versus $\sigma_Y R_N^\ast$ (see text). The line is a linear fit for which R$^2=0.85$ and the prefactor is 2.8.}
\label{fig:modelVolume}
\end{figure}

To refine the interpretation, we rescaled our experimental data with the droplet size. Namely, considering the most compressed state (indicated with an asterisk), the yield stress was multiplied by the neck radius $R_N^{\ast}$. The effective surface tension difference $\Delta\gamma_{LV}$ shows to be proportional to the resulting quantity. The alignment of the data points is better after rescaling (figure \ref{fig:modelVolume}, $\mathrm{R}^2= 0.85$) than for the raw data (figure \ref{fig:V_vs_ys}, $\mathrm{R}^2= 0.74$) and the prefactor is of order 1.

This evidences that even a static surface tension measurement will depend on the flow history, and this all the more as the yield stress is high and the droplet is large. The error on the measurement, if it is performed after (or during) a flow in always the same direction, will be of the order of $\sigma_Y\times r$ with $r$ a dimension of the system.

The length $r$ must be thoroughly identified. In our experiments the bridge radius at the most compressed state $R_N^\ast$ is the characteristic length scale because it corresponds to the greatest force difference in a force-$L$ plot and thus determines $\Delta\gamma_{LV}$.

\subsection{Influence of elasticity}
Results of figure \ref{fig:elasticexp} clearly show that $G'$ has a strong effect on the shape of the stretching-compression cycle. Indeed the elastoplastic force depends on the elastic deformation of the bridge (see part \ref{par:model}), and especially at changes of deformation direction.

\begin{figure}[floatfix,h]
\centering
\includegraphics[width=0.7\columnwidth]{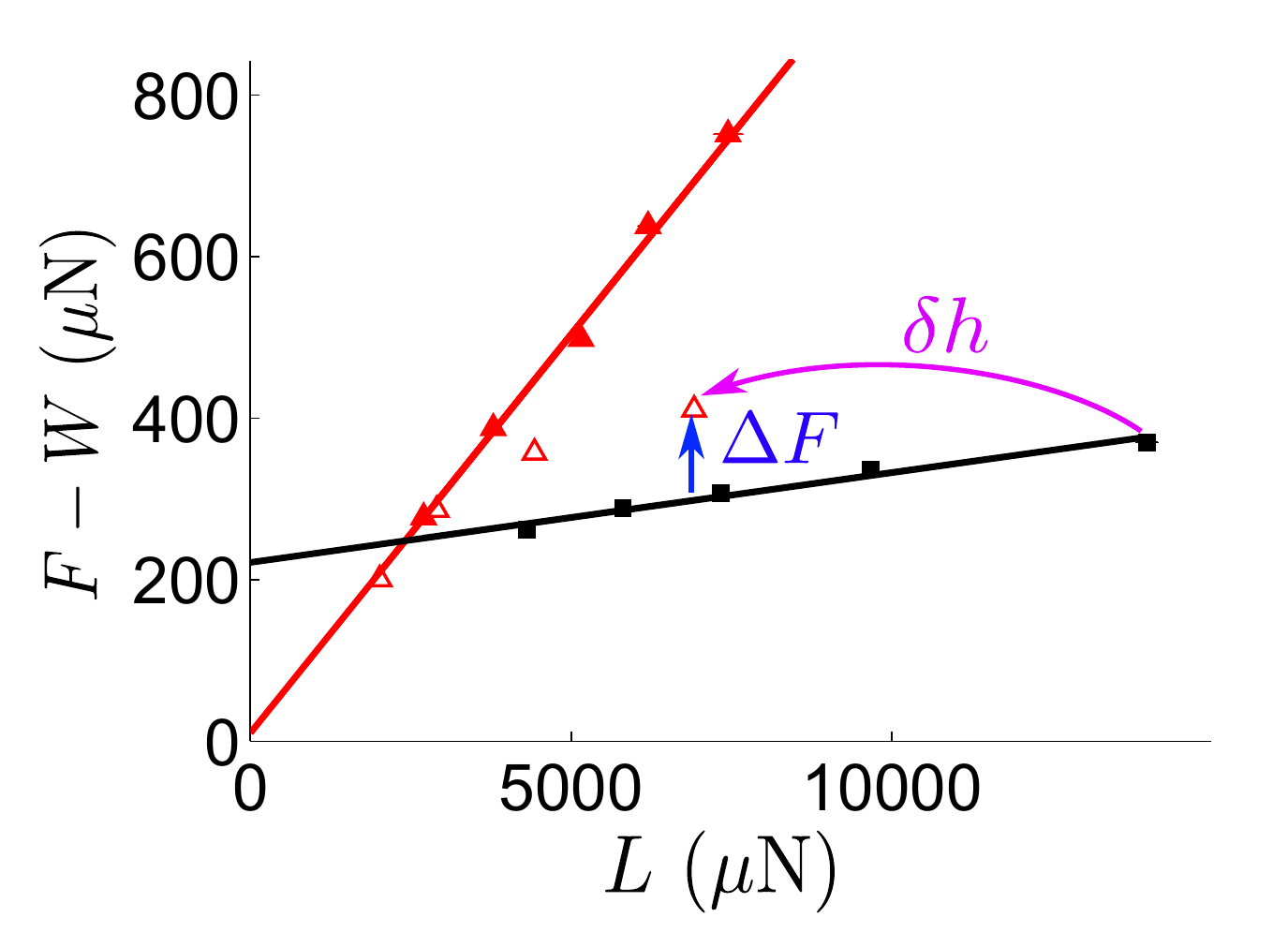} \put(-130,110){a)}\\ \includegraphics[width=0.9\columnwidth]{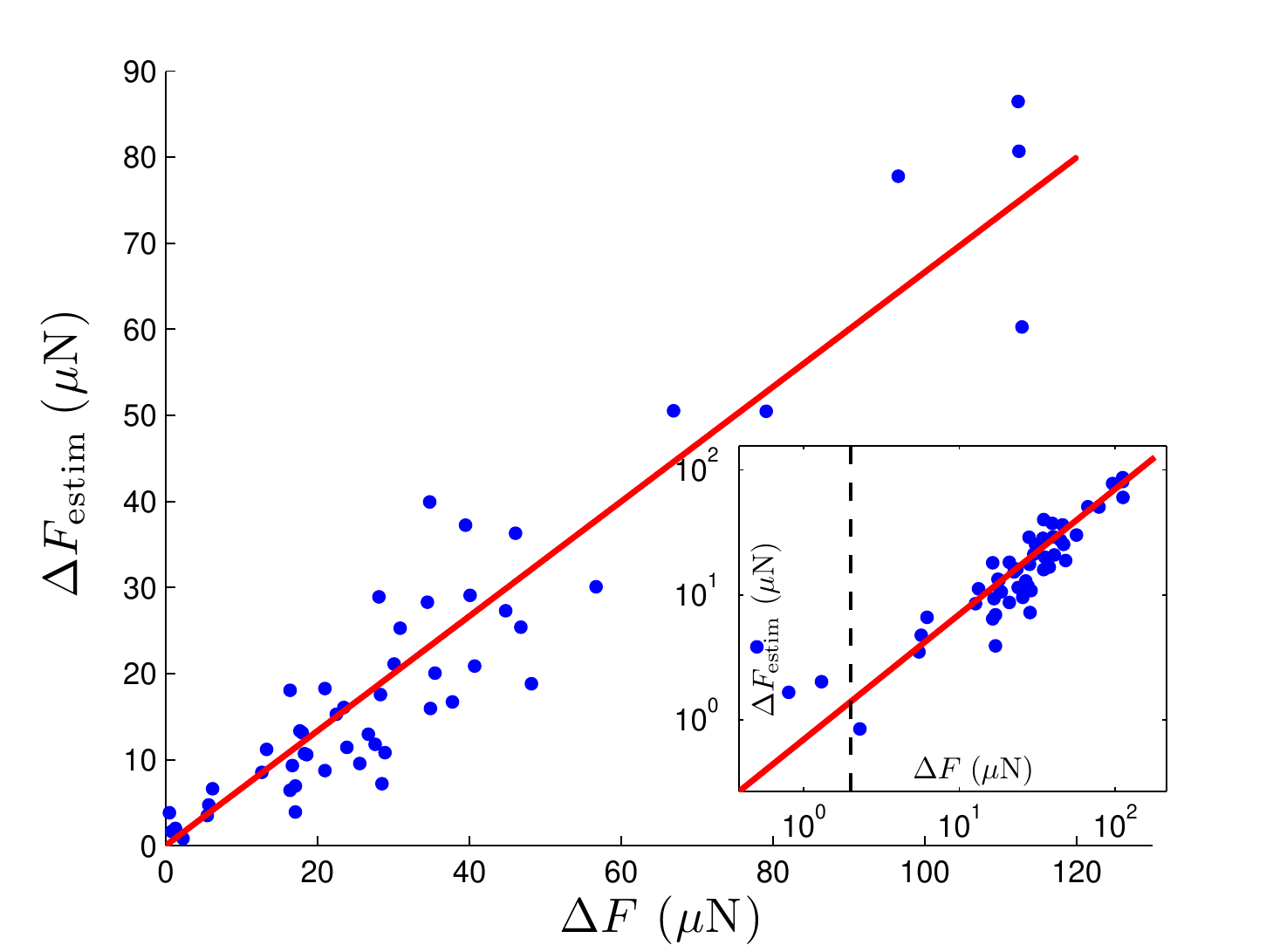} \put(-185,135){b)} 
  \caption{a) Definition of $\Delta F$ in a force-$L$ plot. b) Calculated elastic force (see text) $\Delta F_\text{estim}$ versus $\Delta F$ (measured). The red line is a linear fit with a slope of 0.67. Inset : same plot in log log scale. The dashed line indicates the measurement error on the x-axis. On the y-axis, the error is around 25\% of the values.}
  \label{fig:elastic}
\end{figure}

To analyse this effect in a more systematic way, the difference between the force $F-W$ of the first step of the second stretching and the force corresponding to a compressed bridge at the same $L$ (see figure \ref{fig:elastic}a) was measured on each force-$L$ plot. This force difference $\Delta F$ is plotted as a function of an estimated elastic force $\Delta F_\text{estim}$ in figure \ref{fig:elastic}b. In the filament approximation this elastic force corresponds to:
\begin{equation*}
\Delta F_\text{estim}= 3G' \times \dfrac{\delta h}h \times \pi R_N^2
\end{equation*}
where $3G'$ is an estimation of the Young modulus of the gel, $\frac{\delta h}h$ is the relative variation of the bridge height $h$ on the first step of the second stretching and $\pi R_N^2$ is the section of the bridge at the neck after the first step of the second stretching.

These two forces are nearly equal, which evidences the elastic behavior at the change between compression and stretching. This confirms that the shape of the stretching-compression cycle is driven by the elasticity modulus $G'$.

\subsection{Effect of the initial stress}
Many experimental force-$L$ plots show a positive y-intercept for the stretching part (red fits), whereas this feature does not appear in the model, where this y-intercept is always negative. This can be explained as follows: the model assumes that the initial stress is maximum (ie. the fluid has reached the yielding point) before the first stretching phase. But this cannot be checked experimentally. It is likely that in some experiments the initial stress is not maximum, although the fluid is strongly stretched before the beginning of the experiment.

\begin{figure}[floatfix,h]
\centering
\includegraphics[width=0.9\columnwidth]{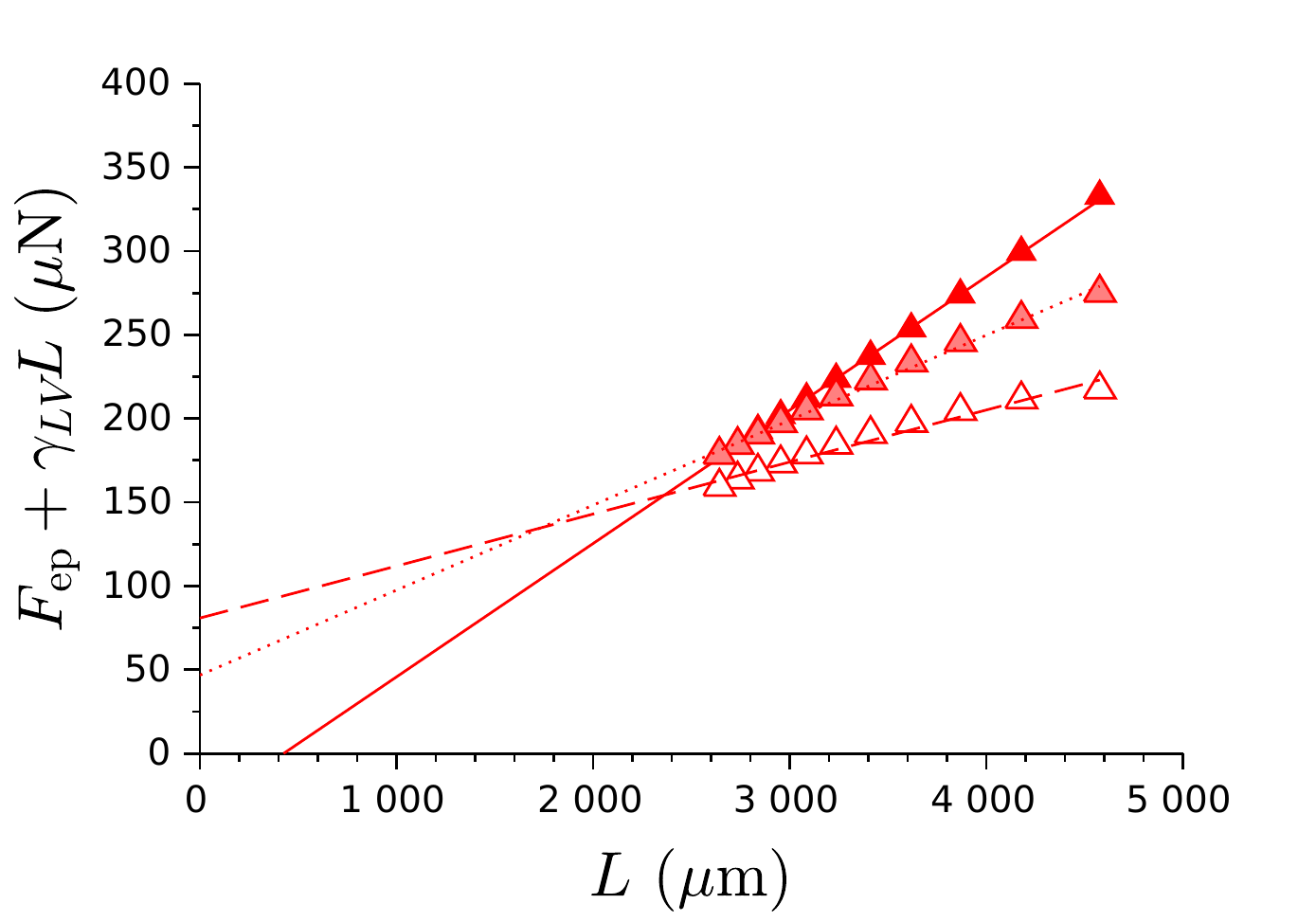}
  \caption{Stretching phase in the filament geometry, with $G'/\sigma_Y=0.5$ and three different initial conditions. Full red symbols: maximum initial stress $\sigma_0=\sqrt3 \sigma_Y$. Pink symbols: zero initial stress $\sigma_0=0$. Empty symbols: negative initial stress $\sigma_0=-\sqrt3\sigma_Y$ (after full compression). Each set of points is shown with its linear fit.}
  \label{fig:initial}
\end{figure}

Three different cases are illustrated in figure \ref{fig:initial} which shows results from the model with different initial conditions: the full red symbols stand for a stretching phase beginning with a maximum stress, the pink symbols for the same stretching series with a zero initial stress and the empty symbols for a stretching phase with maximum negative stress (after a full compression for example). The ratio $G'/\sigma_Y$ is set to 0.5 in this figure and the geometry is filament-like. In the case of a zero initial stress it is clear that even if the points seem nearly aligned, a positive y-intercept arises and the slope $\gamma_{LV}^{\,u}$ decreases.

The model shows that above a ratio $G'/\sigma_Y\approx 8$, the force-$L$ curves are not sensitive any more to the initial stress (see figure \ref{fig:filamentsigma} for example) and the y-intercept of the stretching branch (red) is always negative.

\begin{figure}[floatfix,h]
\centering
\includegraphics[width=0.9\columnwidth]{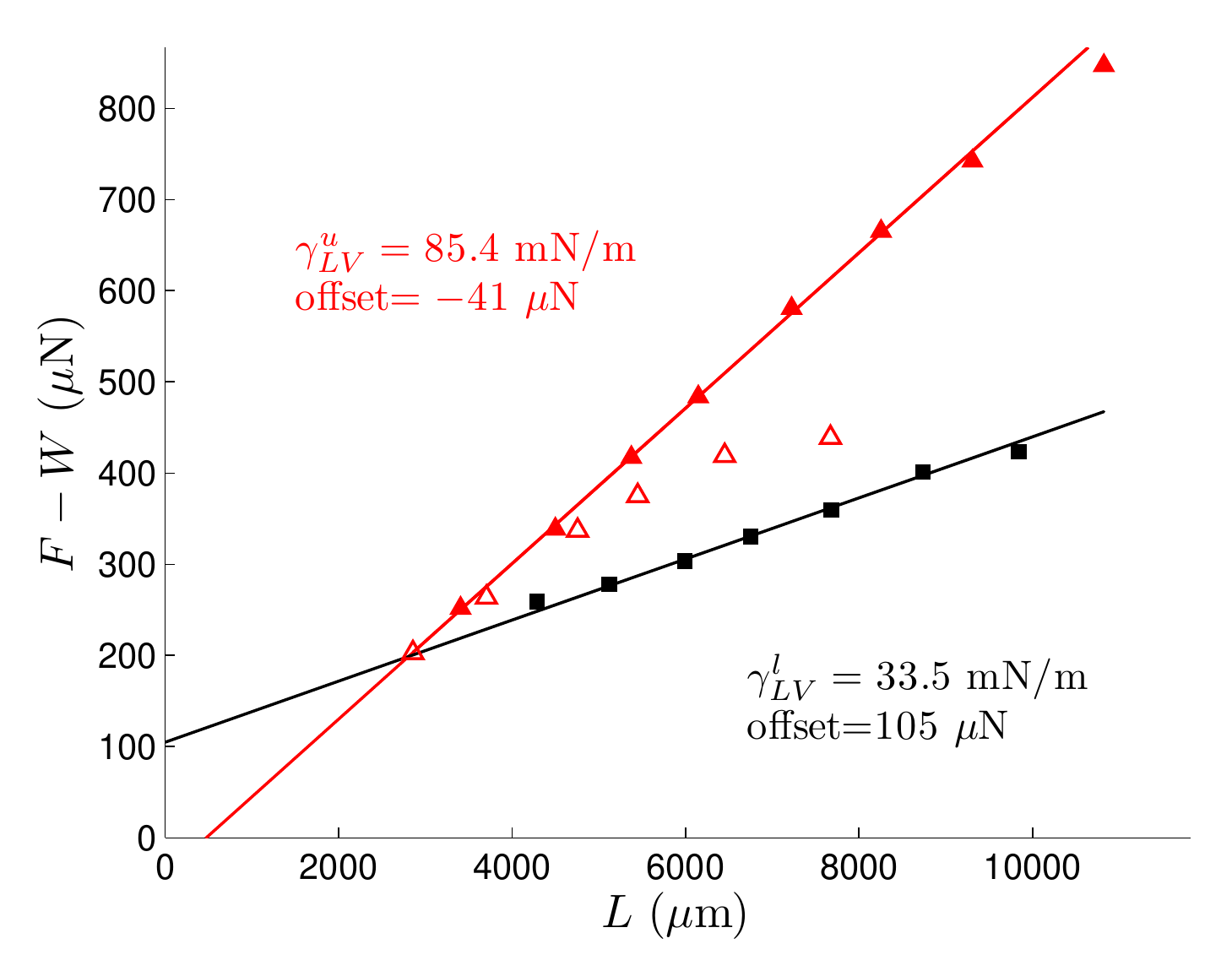}
  \caption{Force-$L$ plot for carbopol 0.75\% (HS) with $G'/\sigma_Y=5.1$. From the second stretching branch (empty triangles) we see that the initial condition is quickly forgotten. The mean slope is 59.5 mN/m. }
  \label{fig:highGprim}
\end{figure}

For the carbopol samples we used, the ratio $G'/\sigma_Y$  is comprised between 2.3 and 6.1. Therefore the curves are sensitive to the initial conditions. This explains the occurrence of positive y-intercepts for the stretching branch and also the dispersion in the limiting slopes values. However for our samples where $G'/\sigma_Y>5$ (HS carbopols), we could obtain force-$L$ plots with negative y-intercept and limited influence of initial stress, as predicted by the elastoplastic model (figure \ref{fig:highGprim}).

\subsection{Surface tension}

\paragraph*{Estimation with the bridge tensiometer.~~} As can be seen in figure \ref{fig:V_vs_ys}a, for vanishing yield stress the upper and the lower slopes both tend to around 63 mN/m. This suggests that the surface tension of carbopol gels is close to this value. For higher values of the yield stress, our experiment clearly shows that the way an experiment is performed (ie. the fact that the yield stress fluid is stretched or compressed) influences a lot the value of the surface tension found via this experiment.

The bridge tensiometer setup provides an easy way to measure the surface tension of yield stress fluids with $G'\approx 8 \sigma_Y$ or more: on a force-$L$ plot the data align on two limiting curves which are symmetrical with respect to $F-W=\gamma_{LV}L$. The true value of the surface tension is thus the mean of the slopes of the two linear fits.  With the model, taking $G'/\sigma_Y=8$ in the filament geometry, the surface tension value could be recovered in this way within less than 1\%. Note that the condition $G'/\sigma_Y>8$ is usually met in a large range of yield stress fluids like emulsions \cite{mason_yielding_1996}, clay suspensions \cite{luu_drop_2009} and microgels pastes \cite{meeker_slip_2004}.

By this method and with our HS samples for which $G'/\sigma_Y>5$ we obtained as a maximal value for the mean slope 63.1 mN/m for 0.25\% carbopol ($\sigma_Y=4.6$ Pa) and 59.5 mN/m for 0.75\% carbopol ($\sigma_Y=15.6$ Pa).

\paragraph*{Ascending bubble measurements.~~} Our results with the bridge tensiometer are confirmed by other experiments with an ascending bubble setup (Teclis Tracker). Here again the apparent surface tension depends on the flow history.

The surface tension between carbopol and air was measured by injecting an air bubble in a large volume of very low yield stress carbopol ($\sim 1$ Pa) and analyzing the bubble profile. The measurement can be static or dynamic.  In the dynamic case, a given volume of air is injected in the fluid (corresponding to a given area $A_0$), and then a fixed interface area is imposed, either greater or smaller than $A_0$. The area remains then fixed thanks to a retroaction loop during the whole measurement, which lasts for about 10 minutes each time.

\begin{figure}[floatfix,h]
\centering
  \includegraphics[width=0.9\columnwidth]{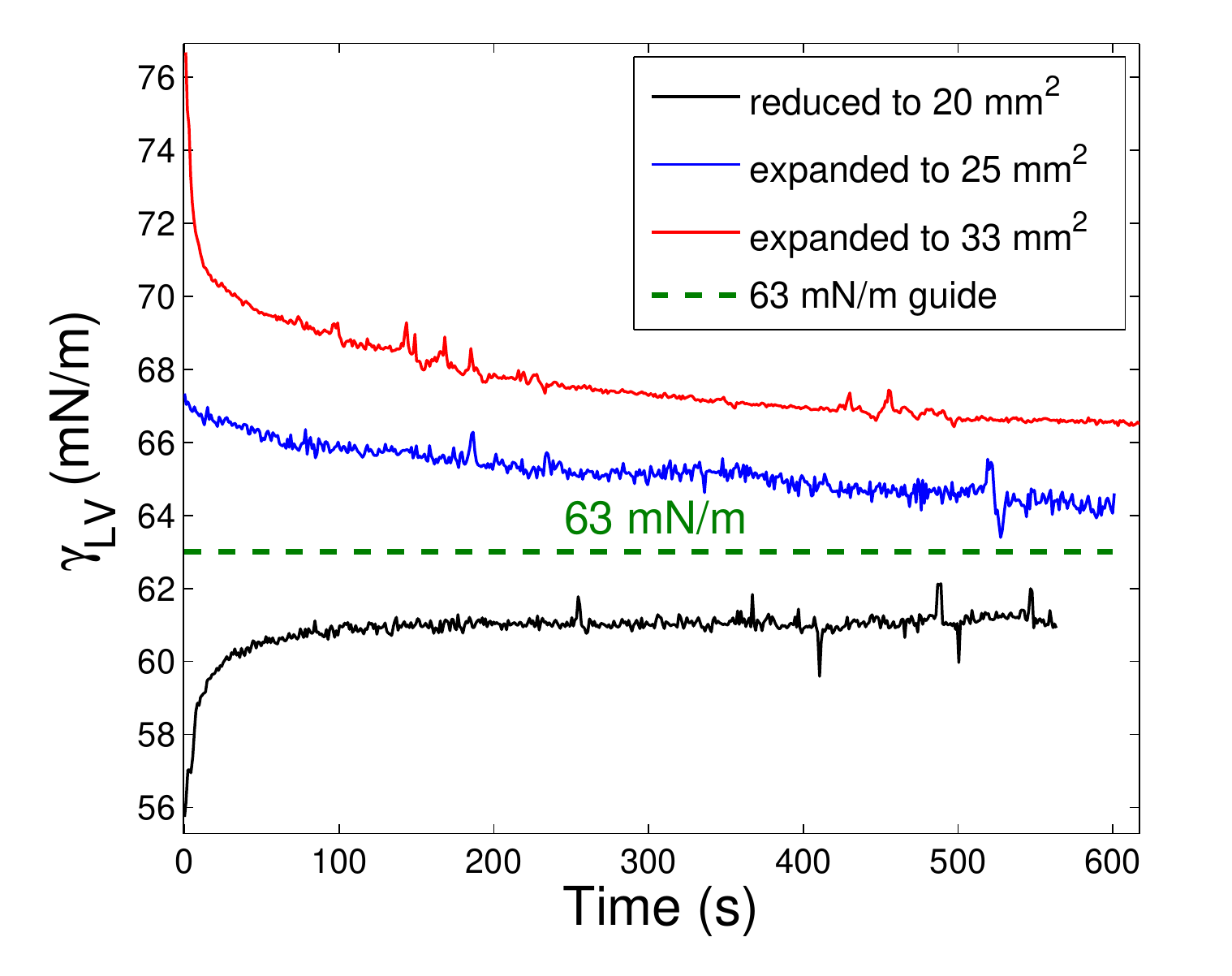}
  \caption{Evolution of surface tension measured by an ascending bubble method, for a 0.25\%(MS) carbopol of yield stress $\sigma_Y=0.3$ Pa, after different changes in interface area at $t=0$. Before $t=0$ the volume of the bubble is 13 $\mu$L and its area is slightly less than 25 mm$^2$.}
  \label{fig:teclis}
\end{figure}

The evolution of the apparent surface tension is plotted in figure \ref{fig:teclis}. A slow relaxation can be observed after the increase (resp. decrease) of the area of the bubble, at $t=0$. The end value of the apparent surface tension is rather stable, but depends on the quantity of area added (resp. removed) at the beginning. We interpret this as an elastic effect which cannot relax totally because of the yield stress of the fluid. For a 0.25\%(MS) carbopol of yield stress 0.3 Pa, the apparent surface tension after 10 minutes is between 61 mN/m and 66 mN/m, depending on the history of the bubble. For a 0.5\%(MS) carbopol of yield stress 1.75 Pa, the apparent surface tension is between 59 mN/m and 65 mN/m.

The ascending bubble commercial device can however not be used for determining the surface tension of a wide range of carbopols, since it is not powerful enough to push a bubble in the liquid when the yield stress is over 2 Pa.

\paragraph*{Comparison with values found in the literature.~~} Different values can be found for carbopol surface tension in the literature. First, Hu \emph{et al.} \cite{hu_surface_1991} found by a maximum bubble pressure method that neutralized carbopol from 0.025\% to 0.1\% had the same surface tension as pure water on a wide temperature range. 10 years later, Manglik \emph{et al.} \cite{manglik_dynamic_2001} found by the same method, but using dynamic and static measurements, that equilibrium surface tension of carbopol decreased clearly when concentration increased. For their maximum polymer concentration (0.2\%), the surface tension is measured at 69~mN/m. However their carbopols seem not to be neutralized and the rheological data are unclear. An explanation for the discrepancy between these references' results and ours could be that the yield stress and elasticity effects were not taken into account, even if Manglik \emph{et al.} corrected their dynamic measurements with a viscosity term.

More recently, Boujlel and Coussot \cite{boujlel_measuring_2013} used a plate withdrawal method to study the effect of the capillary Bingham number $B_c=\sigma_Y\times D/\gamma_{LV}$ on the withdrawal force, changing the yield stress and the dimension $D$ of the plate. An extrapolation to $B_c=0$ allowed them to estimate the surface tension of carbopol to 66~mN/m.

G\'eraud \emph{et al.}\cite{geraud_capillary_2014} performed capillary rises of neutralized carbopol gels and could extract a value for the surface tension from the maximum rise height as a function of the gap. They found $\gamma_{LV}\cos\theta_0=48\pm3$ mN/m, which cannot be compatible with Boujlel's results unless the contact angle $\theta_0$ is at least $43^\circ$. Yet the contact angle of carbopol droplets freshly dropped on the same surfaces used by G\'eraud \emph{et al.} was always measured smaller than $25^\circ$. This value would imply a maximal value of 53 mN/m for the surface tension of carbopol.

Our work is able to reconcile these different measurements, as it shows that the yield stress and elasticity of a fluid give rise to a separation in two different values for the effective surface tension, depending on the flow history. The maximum bubble pressure method and the plate withdrawal method being always performed in surface extension, the energy cost comes from surface increase as well as from the rheological resistance of the gel. We thus expect a higher value than for capillary rises where the fluid is moving forward driven by surface tension.

\section{Conclusions and perspectives}
In this work we have shown that the measurement of a yield stress fluid surface tension is strongly influenced by the fluid rheology and the protocol. In particular, the direction of the flow during or even before the measurement gives rise to excess forces, either positive or negative, that are difficult to compute in a general case. It is thus useful to perform measurements with a setup allowing to test different flow histories.

We have proposed a method to measure the surface tension of yield stress fluids of high elastic modulus ($G'>8\sigma_Y$). The surface tension value is given by the mean of the two limiting slopes in a force-$L$ plot.

To go further, capillary bridges are a convenient tool to study the surface tension and the rheology of simple or complex fluids. Our method can be very precise when it comes to measure the surface tension of simple fluids. With complex fluids, the complexity of the real shape of the bridge does not change qualitatively the adhesion force compared to model geometries.

Future work will focus on the influence of wall slip and line pinning on the adhesion force of capillary bridges and more generally on the measurement of surface tension. In this scope, other experiments, including surface fluctuation spectra analysis \cite{pottier_high_2011}, have already started.

\balance

\bibliography{mybibfile} 
\bibliographystyle{rsc}

\normalsize
\section*{Acknowledgements}
The authors would like to thank V. Bergeron and S. Santucci for letting them use the Teclis Tracker, L. Talini for fruitful discussions and SFSA experiments, and the Institut Universitaire de France (IUF) for funding.

\section*{Appendix A: Typical force evolution during an experiment}
\begin{figure}[floatfix,h]
\centering
  \includegraphics[width=0.9\columnwidth]{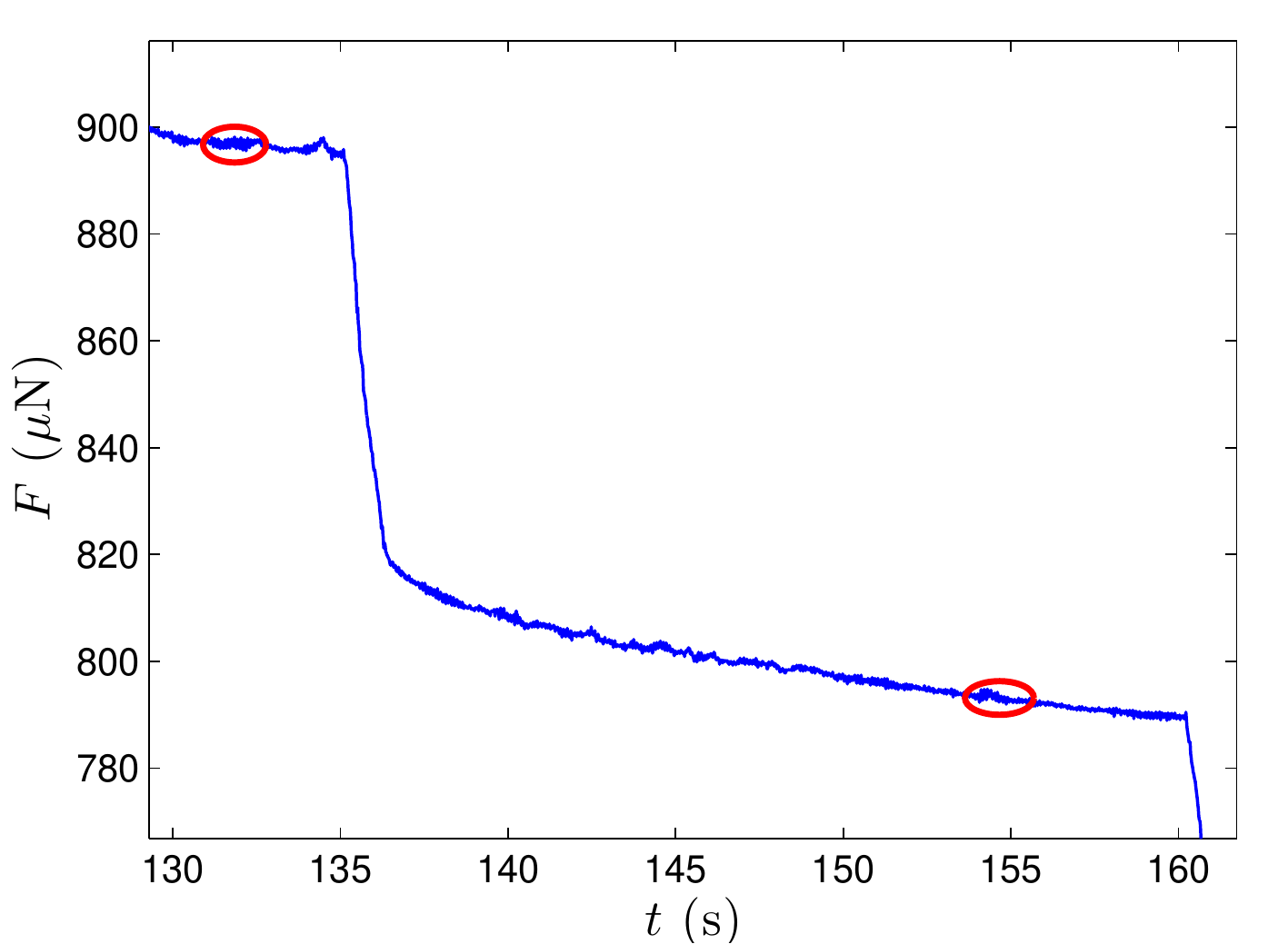}
  \caption{Force evolution in time during a typical stretching step. The red ellipses represent the moments when the force value and the picture are saved.}
  \label{fig:force}
\end{figure}

The evaporation of water and carbopol is not negligible during a whole experiment (about 30 minutes). This is reflected in the force evolution in time even in the absence of flow (figure \ref{fig:force}). The force decreases slowly, at an approximative rate of 1 $\mu$N per second.

To minimize the uncertainties on the force, its value is saved only when this slow evolution is reached and the picture is taken at the same moment. The resulting uncertainty is very small compared to the force values (100 to 1000 $\mu$N) and it is taken into account when fitting the force-$L$ plots.

\section*{Appendix B: Calculation details in the pancake geometry}
The differences between the filament and the pancake geometries lie in the expression of the geometrical parameter $L$, the dominating terms of the stress tensor and then the calculation of the normal force.

We recall the approximate expression of $L$ in a quasi-cylindrical geometry where the volume is $V=\pi R_N^2 h$:
$$L\approx \pi  \sqrt{\dfrac{V}{\pi h}} + \dfrac{2 V \cos \theta_0}{h^2}$$

In the case of a very flat drop, the dominant curvature is in the $(r,z)$ plane so that $L \approx 2 V \cos \theta_0 / h^2$.

In addition the deformations and dissipation are dominantly due to shear along the $z$ direction, in the lubrication approximation. In this case, we cannot use a uniform description like in the filament geometry, but we need to describe the stress profile at the wall across the axis $\sigma_\text{wall}(r) = \sigma_{rz}(r,z = z_\text{wall})$.
Note that the region of high stress in the pancake geometry is located at the walls whereas the highest stress region is located in the neck of the filament.

The extra pressure due to the elastoplastic flow is here denoted $p = \Delta p-\Delta p_\text{Laplace}$. The stress balance reads:
$$
 \frac{\partial \sigma_{rz}}{\partial z}= \frac{d p}{d r}
$$
which is a function of $r$ only, so that $\dfrac{d p}{d r} = \dfrac{2 \sigma_\text{wall}}{h}$ . This allows for the determination of the elastoplastic traction force:
\begin{align*}
F_\text{ep} &= - \int_0^{R_N} 2 \pi r p(r) dr\\ &= \int_0^{R_N} \pi r^2 \frac{d p}{d r} dr\\ &= \int_0^{R_N} \pi r^2  \frac{2 \sigma_\text{wall}}{h} dr
\end{align*}

Each step starts with a given stress profile $\sigma_0(r)$. For a variation in height $\Delta h$, the corresponding step in elastic stress is $\Delta \sigma (r) = 3 G' \Delta h \dfrac r {h^2}$. For each position, we thus evaluate the new stress value at the wall:
$$\sigma_\text{wall}(r)  = \Bigg\{
  \begin{tabular}{lcl}
   $- \sigma_Y$ & \ & if $ \sigma_0(r) + \Delta \sigma (r) < - \sigma_Y $ \\
  $ \sigma_0(r) + \Delta \sigma (r) $ & \ & if $- \sigma_Y  <   \sigma_0(r) + \Delta \sigma (r)  < \sigma_Y $ \\
 $+\sigma_Y$ & \ & if $ \sigma_0(r) + \Delta \sigma (r)  > + \sigma_Y $ \\
  \end{tabular}
  $$
We also re-evaluate the values of the radial positions $r' = r - \dfrac{r \Delta h} {2h}$ and finally evaluate at each step the traction force $$F_\text{ep}  = \int_0^{R_N}   \pi r^2 \frac{2 \sigma_\text{wall}(r)}{h} dr $$

In practice the drop is assumed to be initially stretched just at the onset of yielding so the initial stress is set at $\sigma_0(r) = \sigma_Y (r/R_N)$ (this linear variation corresponds to a pure elastic deformation). Finally we stretch then compress and finally stretch again the drop from $h=0.5$~mm to $h=1.5$~mm by successive steps $\Delta h = 0.1$~mm. For each step,  the total traction force is the sum of the capillary force and the elastic one. Typical curves of traction forces as a function of $L$ are plotted in figure \ref{fig:pancake}. The same phenomenology is observed in the pancake and filament geometries: stretching (resp. compression) of the capillary bridge is associated with an increase (resp. decrease) of the apparent surface tension, which is more pronounced for higher yield stresses. The force cycles also depend on the elastic modulus and initial stress condition. A noticeable difference between the two geometries is the explored range of the geometrical parameter $L$: the pancake geometry corresponds to more compressed drops characterised by higher $L$ values than in the filament geometry.

\end{document}